\let\latexaddtocontents\addtocontents
\documentclass[a4paper,twocolumn,11pt, unpublished]{quantumarticle}
\let\addtocontents\latexaddtocontents

\pdfoutput=1

\usepackage[colorlinks=true,linkcolor=blue,urlcolor=blue,citecolor=blue,anchorcolor=green,pdfusetitle]{hyperref}
\usepackage{amsmath}
\usepackage{verbatim}
\usepackage{graphicx}
\usepackage{xcolor}
\usepackage{bbold}
\usepackage{amssymb}
\usepackage{amsthm}
\usepackage[braket,qm]{qcircuit}
\usepackage{physics}
\usepackage{comment}
\usepackage{tikz}
\usepackage{hyperref}
\usepackage{algorithm}
\usepackage{xspace}
\usepackage[utf8]{inputenc}
\usepackage{algpseudocode}
\usepackage[inkscapelatex=false]{svg}
\usepackage[style=numeric-comp,
backend=bibtex8,
doi=true,
isbn=false,
url=false,
eprint=false,
maxbibnames=5,
sorting=none]{biblatex}
\theoremstyle{plain}

\begin{document}

\title{
qLUE: A Quantum Clustering Algorithm for Multi-Dimensional Datasets
}
\author{Dhruv Gopalakrishnan}
    \email{dgopalak@uwaterloo.ca}
    % \thanks{this author contributed equally}
    %\thanks{These authors contributed equally}
    %\email{ariel.shlosberg@colorado.edu}
    %\email{ajjena@uwaterloo.ca}
    %\affiliation{Department of Physics \& Astronomy, University of Waterloo, Waterloo, ON N2L 3G1, Canada}
    \affiliation{Institute for Quantum Computing, University of Waterloo, Waterloo, ON N2L 3G1, Canada}
    \affiliation{Department of Electrical and Computer Engineering, University of Waterloo, Waterloo, ON N2L 3G1, Canada}
    \affiliation{Perimeter Institute of Theoretical Physics, Waterloo, ON N2L 2Y5, Canada}
\author{Luca Dellantonio}
    \email{l.dellantonio@exeter.ac.uk}% Your name
    % \thanks{this author contributed equally}
    \affiliation{Institute for Quantum Computing, University of Waterloo, Waterloo, ON N2L 3G1, Canada}
    \affiliation{Department of Physics \& Astronomy, University of Waterloo, Waterloo, ON N2L 3G1, Canada}
    \affiliation{Department of Physics and Astronomy, University of Exeter, Stocker Road, Exeter EX4 4QL, United Kingdom}
\author{Antonio Di Pilato}
    \email{tony.dipilato@cern.ch}
    % \thanks{this author contributed equally}
    %\thanks{These authors contributed equally}
    %\email{ariel.shlosberg@colorado.edu}
    %\email{ajjena@uwaterloo.ca}
    \affiliation{CERN, Geneva}
\author{Wahid Redjeb}
    \email{wahid.redjeb@cern.ch}
    % \thanks{this author contributed equally}
    %\thanks{These authors contributed equally}
    %\email{ariel.shlosberg@colorado.edu}
    %\email{ajjena@uwaterloo.ca}
    \affiliation{CERN, Geneva}
    \affiliation{RWTH Aachen University Physikalisches Institut III A, Aachen, Germany}
\author{Felice Pantaleo}
    \email{felice.pantaleo@cern.ch}
    % \thanks{this author contributed equally}
    %\thanks{These authors contributed equally}
    %\email{ariel.shlosberg@colorado.edu}
    %\email{ajjena@uwaterloo.ca}
    \affiliation{CERN, Geneva}
\author{Michele Mosca}
    \email{michele.mosca@uwaterloo.ca}
    % \thanks{this author contributed equally}
    %\thanks{These authors contributed equally}
    %\email{ariel.shlosberg@colorado.edu}
    %\email{ajjena@uwaterloo.ca}
    %\affiliation{Department of Physics \& Astronomy, University of Waterloo, Waterloo, ON N2L 3G1, Canada}
    \affiliation{Institute for Quantum Computing, University of Waterloo, Waterloo, ON N2L 3G1, Canada}
    \affiliation{Perimeter Institute of Theoretical Physics, Waterloo, ON N2L 2Y5, Canada}
    \affiliation{Department of Physics \& Astronomy, University of Waterloo, Waterloo, ON N2L 3G1, Canada}
    \affiliation{Department of Combinatorics and Optimization, University of Waterloo, Waterloo, ON N2L 3G1, Canada}
\begin{abstract}
Clustering algorithms are at the basis of several technological applications, and are fueling the development of rapidly evolving fields such as machine learning. In the recent past, however, it has become apparent that they face challenges stemming from datasets that span more spatial dimensions. In fact, the best-performing clustering algorithms scale linearly in the number of points, but quadratically with respect to the local density of points. In this work, we introduce qLUE, a quantum clustering algorithm that scales linearly in both the number of points and their density. qLUE is inspired by CLUE, an algorithm developed to address the challenging time and memory budgets of Event Reconstruction (ER) in future High-Energy Physics experiments.
As such, qLUE marries decades of development with the quadratic speedup provided by quantum computers. We numerically test qLUE in several scenarios, demonstrating its effectiveness and proving it to be a promising route to handle complex data analysis tasks -- especially in high-dimensional datasets with high densities of points. The code we used for these simulations is available at Ref.~\cite{github1}
%\href{https://github.com/godspeed5/QLUE}{https://github.com/godspeed5/QLUE}
\end{abstract}
\maketitle
%

%
%\tableofcontents
%

%
\section{Introduction}
\label{sec:introduction}

Clustering is a data analysis technique that is crucial in several fields, owing to its ability to uncover hidden patterns and structures within large datasets. It is essential for simplifying complex data, improving data organization, and enhancing decision-making processes \cite{simplifyenrichment, clusteringapps, decision, decisionreview}. For instance, clustering has been applied in marketing \cite{marketresearch, HUANG2007313}, where it helps segment customers for targeted advertising \cite{targetedads}, and in biology, for classifying genes and identifying protein interactions \cite{Dutta2020, geneclassification, Wang2010, ensembleppinteraction}. In the realm of computer science and artificial intelligence, it is invaluable for image \cite{clusteringsegmentation} and speech recognition \cite{speechclustering, Chang_2017_ICCV}, as well as recommendation systems \cite{clusterrec, theontology} that personalize content for users. Finally, clustering techniques are pivotal for Event Reconstruction (ER), where data points that originated from the same ``event'' must be collected together. In High-Energy Physics, for instance, clustering algorithms reconstruct the trajectories of subatomic particles in collider experiments. It is expected that the endcap high granularity calorimeter (HGCAL) \cite{CMS:2017jpq} being built for the CMS detector at the High Luminosity Large Hadron Collider will provide extremely large volumes of data that must be tackled by new generations of clustering algorithms such as CLUE.  The discovery of the Higgs boson \cite{Aad_2012}, awarded the Nobel prize in 2012, was made possible by such algorithms. 

ER enables the interpretation of data obtained from particle collision events, including those occurring at the Large Hadron Collider (LHC) at CERN. Several clustering algorithms like DBScan, K-Means, and Hierarchical Clustering among others \cite{Amaro_2023, Rodenko_2019, DALITZ2019159} can be employed for ER. Our work is based on CERN's CLUstering of Energy (CLUE) algorithm \cite{clue}, which is adopted by the CMS collaboration \cite{fourtop, betameson, Hayrapetyan:2890630}. It is designed for the future HGCAL detector due to the limitations of the currently employed algorithms. Despite these limitations, such algorithms are already at the basis of several discoveries, such as the doubly charged tetraquark \cite{tetraquark}, the observation of four-top quark production in proton-proton collisions \cite{fourtop} and the study of rare B meson decays to two muons \cite{betameson}.

The efficiency of clustering algorithms, as illustrated by the CLUE algorithm \cite{clue}, is crucial for handling extensive datasets. Initially designed for two-dimensional datasets, CLUE reduces the search complexity from $O(n^2)$ to $O(mn)$ through the use of local density and a tiling procedure, where $n$ ($m$) represents the (average) number of points (per tile). 

\begin{figure}[!htb]
    \centering
    \includegraphics[width=\columnwidth]{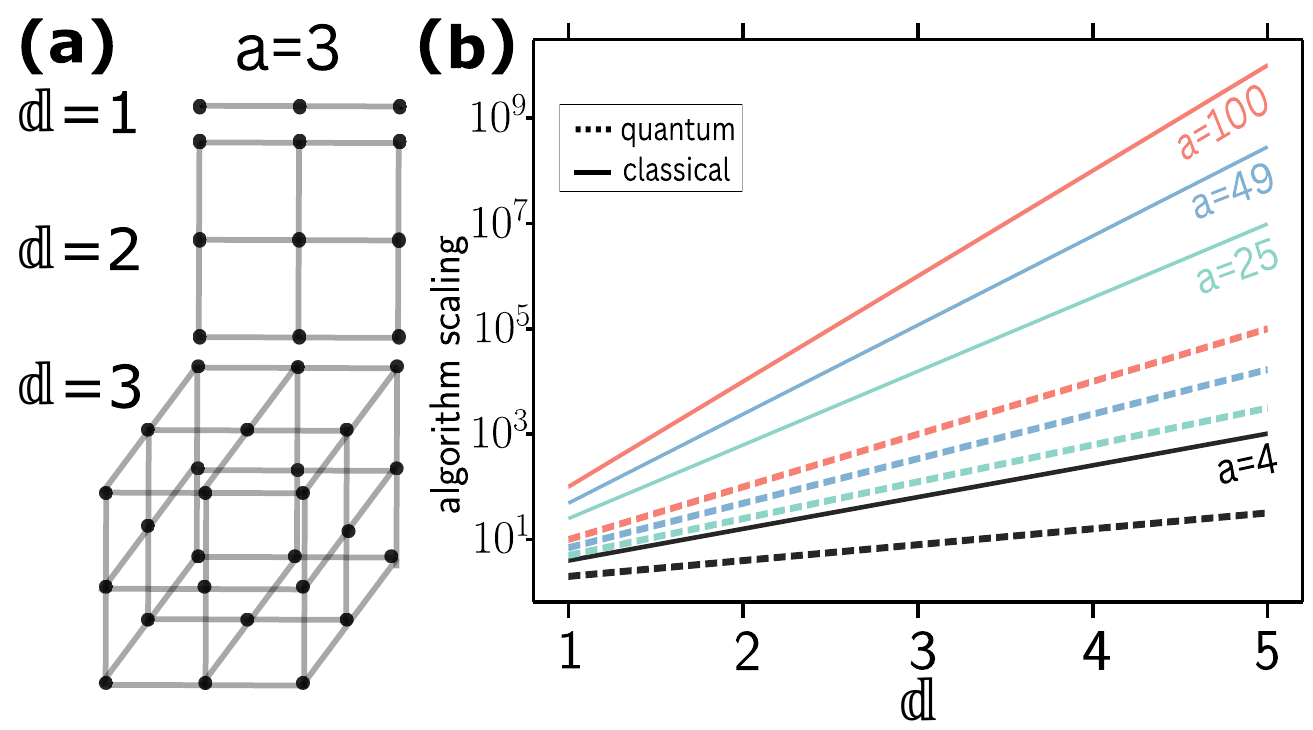}
    \caption{Scaling of point density and complexities of classical and quantum algorithms for the unstructured search problem with dimension $\mathbb{d}$. In (a), different $\mathbb{d}$-dimensional lattices for $\mathbb{d} = 1,2,3$ and $a=3$ points per edge. In (b), best-known classical (solid lines) and quantum (dashed lines) algorithmic scalings for the Unstructured Search Problem \cite{grover} applied to square $\mathbb{d}$-dimensional lattices with the values of $a$ reported in the plot. Classically, the cost $O(m)$ reflects the need to iterate through all the $m$ points to find the desired one. Grover achieves the same in $O(\sqrt{m})$ steps, providing a quadratic advantage. This advantage increases with the density of points in the considered dataset, which grows exponentially with respect to the dimension $\mathbb{d}$ according to $m = a^\mathbb{d}$.}
    \label{fig:intro_pic}
\end{figure}

In the context of CLUE, where data sets are limited to two dimensions, $m$ is small, making this approach to ER particularly effective. However, as the dimensionality of the dataset is incremented, the value of $m$ generally increases exponentially. This is highlighted by Fig.~\ref{fig:intro_pic}(a), where for a $\mathbb{d}$-dimensional lattice with $a$ points per edge, $m$ follows the relation $m = a^\mathbb{d}$. This is a serious challenge to CLUE and classical clustering algorithms in general.

A first step towards extending CLUE to more dimensions is done by 3D-CLUE \cite{clue, clue_tech}. In this work, data points from different layers of detectors are first projected onto a single $\mathbb{d} = 2$ surface, where clustering is then performed.
%In response to the evolving needs of ER tasks, an improved version of CLUE, 3D-CLUE, was introduced \cite{clue, clue_tech}. In this work, it was demonstrated that employing the $\mathbb{d}=3$ features of the data from LHC \cite{clue, clue_tech, cmscollaboration2023development} (rather than restricting to $\mathbb{d} = 2$ sliced sub-datasets as in CLUE) greatly improves the quality of the reconstructed clusters.
However, this projection from the original $\mathbb{d} = 3$ dataset to a $\mathbb{d} = 2$ surface comes at the cost of a slower algorithm since $m$ becomes effectively larger. The solid lines in Fig.~\ref{fig:intro_pic}(b) show the increase in average points per tile in $\mathbb{d}$-dimensional datasets made of the lattices in panel (a). 
%In other words, $m$ is generally larger, following the relation $m = a^3$. 
While the improved performance of 3D-CLUE in ER tasks \cite{clue, clue_tech} justifies the increased computational overhead, extending this enhancement to higher dimensions and larger datasets is challenging.
%, given the exponential dependence of $m$ on dimensionality ($m = a^\mathbb{d}$), 
Finding practical approaches to deal with datasets where $\mathbb{d}$ is large is therefore extremely important, not only for ER tasks but also in other fields such as gene analysis \cite{bioinf} and market segmentation in business \cite{markets}.

Quantum computers provide a route to mitigate the complexity blowup arising from higher dimensional datasets. 
%By exploiting the advantage of Grover algorithm \cite{grover1996fast} and annealing \cit, 
Ref.~\cite{weiharrowthaler} addresses the task of jet clustering in High-Energy Physics, while Ref.~\cite{kerenidis} targets spectral clustering, which itself uses the efficient quantum analog of $k$-means clustering \cite{qmeans}. Other approaches include quantum $k$-medians clustering \cite{quantum_clustering} and a quantum algorithm for density peak clustering \cite{Magano2023}.

%In this work we propose qLUE, a CLUE-inspired, quantum clustering algorithm that exploits quantum computers to mitigate the complexity blowup arising from high dimensional dataset. qLUE  For example, Ref.~\cite{weiharrowthaler} provide a quantum algorithm to speed up the asymptotic complexity of jet clustering in High-Energy Physics by providing quantum annealing and Grover Search-based speedups to the various subroutines. Similarly, Ref.~\cite{kerenidis} provides a quantum speedup to spectral clustering which itself uses the efficient quantum analogue of $k$-means clustering \cite{qmeans}. To leverage the quantum scaling advantage, quantum $k$-medians clustering, divisive clustering and clustering via a neighbourhood graph construction have been developed via the quantization process \cite{quantum_clustering}. In Ref.~\cite{Magano2023}, the authors introduce a quantum algorithm that speeds up the decision version of density peak clustering. Several Quantum methods for solving ER also exist \cite{nicotra2023quantum,T_ys_z_2020}. 

%Building upon the effectiveness of 3D-CLUE and quantum clustering algorithms, we enhance CLUE's capabilities by leveraging the quadratic speedup provided by Grover's algorithm \cite{grover1996fast}.
In this work we develop qLUE, a CLUE-inspired quantum algorithm. 
%for ER tasks. 
Similarly to other quantum algorithms \cite{nicotra2023quantum,T_ys_z_2020}, qLUE leverages the advantage provided by Grover Search \cite{grover}. 
%to mitigate the complexity blow-up associated with high density of points.
%The quadratic speedup provided by Grover Search can mitigate the complexity blow-up associated with high density of points that is expected in high-dimensional datasets. 
A comparison of classical and quantum (Grover) runtimes 
%for the state-of-the-art solutions to the Unstructured Search Problem 
is presented in Fig.~\ref{fig:intro_pic}(b), where the solid [dashed] lines refer to the classical $O(m)$ [quantum $O(\sqrt{m})$] scaling. 
%Different colours refer to the reported values of $a$. 
As can be seen, the complexity advantage that Grover search provides can be substantial, particularly for large values of $\mathbb{d}$ or $a$.
%, where $a$ indicates the number of lattice point per dimension. %For instance, considering $\mathbb{d}=3$ and $a=100$ points per edge, the difference in scaling between the classical and quantum is a factor of $1000$. This factor becomes $10000$ and $100000$ for $\mathbb{d}= 4$ and $\mathbb{d}=5$, respectively.

% An important remark is that while the quantum algorithm may have a more favourable scaling, $O(\sqrt{m}n)$ compared to $O(mn)$, it may be slower to perform individual algorithmic steps compared to its classical counterpart. A question to address is therefore how much slower the quantum computer can be while still being advantageous. This is explained in Fig.~\ref{fig:intro_pic}(b) where we display the maximum overhead $\beta$ for which the quantum algorithm is still more efficient than the classical version. Specifically, $\beta$ here represents all overheads of the quantum device. As it is possible to see from Fig.~\ref{fig:intro_pic}(b), $\beta$ scales as the difference in scaling between classical and quantum algorithms and therefore follows the values mentioned in the previous paragraph, i.e, $1000$ at $d=3$, $10000$ at $d=4$, and $100000$ at $d=5$ for $a=100$.

% We include performance analyses of qLUE simulated on a classical computer in Sec.~\ref{sec:results}. These analyses include noise and overlap characterizations and study of performance on non-centroidal clusters.

Overall, we find that qLUE performs well in a wide range of scenarios. With ER-inspired datasets as a specific example, we demonstrate that clusters are correctly reconstructed in typical experimental settings. Similar to other quantum approaches to clustering that rely on Grover Search \cite{quantum_clustering, pires2021digital}, qLUE also showcases a quadratic speedup compared to classical algorithms. The specific advantages of qLUE are its CLUE-inspired approach to cluster reconstruction (which demonstrated to be extremely successful \cite{fourtop, cmscollaboration2024review, betameson}), and its consequent seamless integration with the classical framework currently employed by the CMS collaboration \cite{clue, clue_tech, cmscollaboration2023development}.

This paper is structured as follows. In Sec.~\ref{sec: qLUE}, we describe our algorithm qLUE. Specifically, we provide a general overview of its subroutines -- namely the Compute Local Density, Find Nearest Higher, and the Find Seeds, Outliers and Assign Clusters steps. We describe the results of our simulated version of qLUE on a classical computer in Sec.~\ref{sec:results}. In more detail, we explain the scoring metrics we use to quantify our results, and describe qLUE performance when the dataset is subject to noise and different clusters overlap. Conclusions and outlook are finally presented in Sec.~\ref{sec:conclusions}.

\begin{figure*}[!htb]
    \centering
    \includegraphics[width=\textwidth]{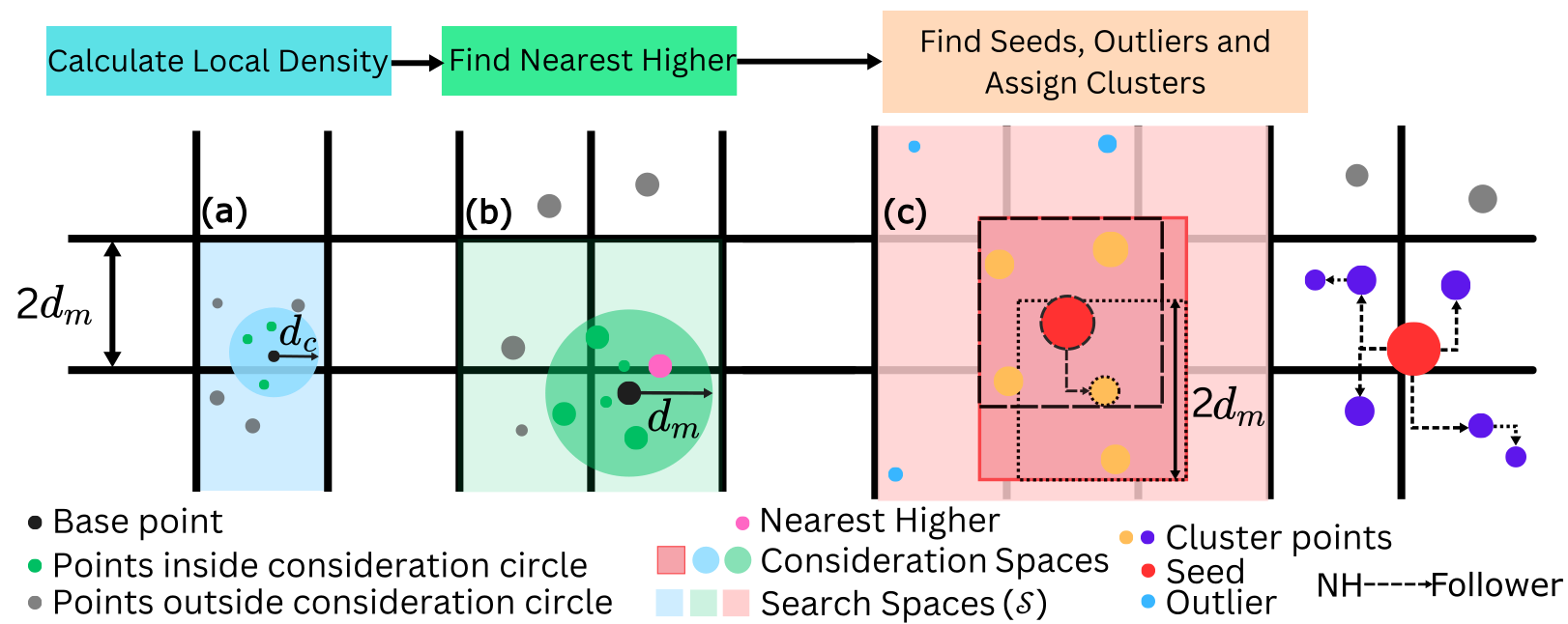}
    \caption{Pictorial representation of the main subroutines of qLUE. In (a), the Local Density computation subroutine is represented. The consideration circle of radius $d_{\rm c}$ (light blue) centered at the base point $j$ (black) contains all points (green) that satisfy $d_{i,j} \leq d_{\rm c}$. This consideration circle intersects $2$ tiles $\square_k$ (indexed by tile index $k$), highlighted in blue, that form the search space $\mathcal{S}$. As per Eq.~\eqref{eq: LE_computation}, the Local Density computation step determines the set of green points from all points in the search space (green and grey) and then computes the local density. In (b), we pictorially present the Find Nearest Higher ($\mathcal{N}_j$) subroutine. The consideration circle (green) around base point $j$ (black) has radius $d_{\rm m}$. This consideration circle, containing the green points as well as the Nearest Higher $\mathcal{N}_j$ (pink), intersects the $4$ tiles highlighted in green, which form the search space $\mathcal{S}$. 
    %$\mathcal{N}_j$ is determined by qLUE's Nearest Higher computation subroutine whose flow is diagrammatically represented in Fig.~\ref{fig:nh_flow} and described in detail in Sec.~\ref{subsec: nh}. 
    In (c), we describe the Find Seeds, Outliers and Assign Clusters subroutines. The seeds (red) and outliers (blue) are determined via Grover search on the dataset. In this specific example there are two clusters in the dataset whose non-seed points are in orange and purple, respectively. Followers (see main text) in these clusters are connected by dashed arrows. The Cluster Assignment subroutine is shown to be working on the orange cluster where the cluster $\mathcal{C}$ currently consists of the seed (red, dashed border) and the first of its followers (orange, dotted border). Followers are being found within the Dynamic Search Space (DSS, light red box with solid red border). The DSS is formed as the set of tiles $\square_k$ covered partially or fully by the minimum bounding box of the square windows that contains all the search spaces $\mathcal{S}$ of the points within $\mathcal{C}$.
    %determined by the points in $\mathcal{C}$. To determine the DSS, we first open square windows of size $2d_{\rm m}$ centered at the points in $\mathcal{C}$ (the squares with the dashed and dotted borders around the seed and its marked follower respectively). The DSS is formed as the set of tiles intersected by the minimum bounding box (light red box with solid red border) of the square windows. Since it touches the set of nine tiles highlighted in red, this set of tiles is the search space DSS.
    }
    \label{fig:full_algo}
\end{figure*}
\section{qLUE}
\label{sec: qLUE}
qLUE is a quantum adaptation of CERN's CLUE and 3D-CLUE algorithms \cite{clue, clue_tech}, that is specifically developed for ER, yet it is suitable to work with any (high dimensional) dataset. The main advantage of qLUE stems from employing Grover's algorithm, which provides a quadratic speedup for the Unstructured Search Problem \cite{grover}. While qLUE is designed to work in arbitrary dimensions, for clarity we restrict ourselves to $\mathbb{d}=2$. This simplifies the following discussions and allows us to simulate qLUE with meaningful datasets on a classical computer. Generalizations to higher dimensions can be done following the steps outlined below. Furthermore, to provide a better connection with CLUE and 3D-CLUE, we employ a similar notation. 

In Sec.~\ref{subsec: overview}, we offer an overview of the algorithm and its different subroutines. Sec.~\ref{subsec: local} is dedicated to the first subroutine of qLUE, namely, calculating the Local Density. We then explain how to determine the Nearest Highers ($\mathcal{N}_j$), Seeds, and Outliers in Sec.~\ref{subsec: nh}. Finally, Sec.~\ref{subsec: cluster} delves into the conclusive Cluster Assignment subroutine, where the points in the dataset are effectively heirarchically clustered.

\subsection{Overview and Setting}
\label{subsec: overview}
As for CLUE and 3D-CLUE \cite{clue, clue_tech}, we consider a dataset with spatial coordinates and an energy for every point. Similar datasets can also be found in medical image analysis and segmentation \cite{med_image, watershed}, in the analysis of asteroid reflectance spectra   and hyperspectral astronomical imagery in astrophysics \cite{astrophysics, GAFFEY2010564, Gao_2021_CVPR} and in gene analysis in bioinformatics \cite{bioinf, bioinfo}.

In $\mathbb{d}=2$ dimensions, the spatial coordinates $X_j$ for point $j$ are $X_j = \left[ x_{j,1} , x_{j,2} \right]$, that are promptly generalized for larger values of $\mathbb{d}$. Both CLUE and qLUE first perform tiling over the dataset to reduce the search and therefore enhance the efficiency of the algorithm. Tiling is the process of partitioning the dataset into a grid of rectangular tiles $\square_k$, where $k$ is the tile index (see Fig.~\ref{fig:full_algo}). Therefore, our input dataset comprises of point and tile indices $j$ and $k$, respectively, the coordinates $X_j$, and a parameter $E_j$ associated to each point. Following CLUE's notation, we call $E_j$ the energy, yet this should be considered as a label that can be employed to improve the clustering quality for any given dataset. The tiling procedure of qLUE and CLUE enables searching only over Search Spaces $\mathcal{S}$ marked by the tiles in green in Fig.~\ref{fig:full_algo}(a) as opposed to the full dataset. In case of CLUE, this allowed for an improvement in scaling from $O(n^2)$ to $O(m n)$. The scaling of qLUE is investigated below.

%In the case of CERN's CLUE, this allowed for enhanced performances in ER tasks \cite{clue}, while we show in Sec.~\ref{sec:results} that it is also beneficial for non-centroidal clusters.

In this work, we employ a qRAM to store and access data, which is an essential building block for quantum computers. Following Ref.~\cite{Giovannetti_2008}, we therefore assume that we can efficiently prepare the state
\begin{equation}
\label{eq: dataset}
 \sum_j \ket{j} \xrightarrow{\text{qRAM}} \ket{j} \ket{D_j},
\end{equation}
where $D_j$ is the data associated with a given index $j$, e.g. the $j^{\rm th}$ point in the database. For convenience, here and throughout this paper we do not explicitly write the normalization factors of quantum states. 

The qLUE algorithm consists of the following steps:
\\
\underline{\textit{Local Density}}: The first step is to calculate the local density $\rho_j$ of all points $j$ [e.g, black point in Fig.~\ref{fig:full_algo}(a)] that is defined by
\begin{equation}
    \rho_j =  E_{j} + \frac{1}{2}\sum_{d_{i,j} < d_{\rm c}}E_i
    \label{eq: LE_computation}
\end{equation}
and it is indicative of the energy in a neighbourhood of point $j$. As can be seen from Eq.~\eqref{eq: LE_computation} and Fig.~\ref{fig:full_algo}(a), $\rho_j$ is a weighted sum over the energies $E_i$ of all points $i$ whose distance $d_{i,j}  = \sqrt{ \sum_{\alpha = 1}^{\mathbb{d}} (x_{i,\alpha}-x_{j,\alpha})^2  }$ from the base point $j$ is within a user-specified critical radius $d_{\rm c}$ that characterizes the consideration circle for the Local Density computation subroutine (light blue circle in the figure).
As such, $E_i$ is the energy of the $i^{\rm th}$ point which is $d_{i,j}$ away from point $j$. The choice of weight $1/2$ for $E_j$ in the definition of $\rho_j$ in Eq.~\eqref{eq: LE_computation} is empirically found to yield better performances for CLUE \cite{clue}.
\\
\underline{\textit{Find Nearest Higher}}: After calculating the local densities, we determine the nearest highers. The Nearest Higher $\mathcal{N}_j$ of a point $j$ is the point nearest to $j$ with a higher local density $\rho_{\mathcal{N}_j} > \rho_j$. As better explained in Sec.~\ref{subsec: cluster}, the Nearest Higher are used to heirarchically cluster points together in the Cluster Assignment process at the end of qLUE. In Fig.~\ref{fig:full_algo}(b), the Nearest Higher $\mathcal{N}_j$ of the base point $j$ (black point) is the pink point. 
\\
\underline{\textit{Find Seeds, Outliers and Assign Clusters}}: As schematically represented in Fig.~\ref{fig:full_algo}(c), seeds (red points) are the points whose distance $d_{j, \mathcal{N}_j}$ from their Nearest Higher $\mathcal{N}_j$ and whose local density $\rho_j$ are lower bounded by user defined thresholds. Outliers (blue points) are the points whose distance from Nearest Higher is similarly lower bounded but whose Local Density has an upper threshold. As such a point $j$ is 
\begin{subequations} \label{eq: so_conditions}
\begin{align}
    \text{a \emph{seed} if } & d_{\mathcal{N}_j,j} > d_{\rm c} \text{ and } \rho_{j} > \Tilde{\rho} \label{eq:seed}, \\
    \text{an \emph{outlier} if } & d_{\mathcal{N}_j,j} > \delta d_{\rm c} \text{ and } \rho_{j} < \Tilde{\rho} \label{eq:outlier}.
\end{align}
\end{subequations}
Here, $\delta$ is the Outlier Delta Factor that determines the upper bound on the allowed local density for outliers. Furthermore, $\Tilde{\rho}$ is the critical density threshold -- the lowest local density a point can have to be classified as a seed. Both $\delta$ and $\Tilde{\rho}$ are user-specified and can be varied to enhance the quality of the output. Seeds are generally located in areas of high energy density, and will be employed as starting points to build clusters. Outliers are points that are likely to be noise in the dataset and are therefore discarded.
\begin{figure}
    \centering
    \includegraphics[width=\columnwidth]{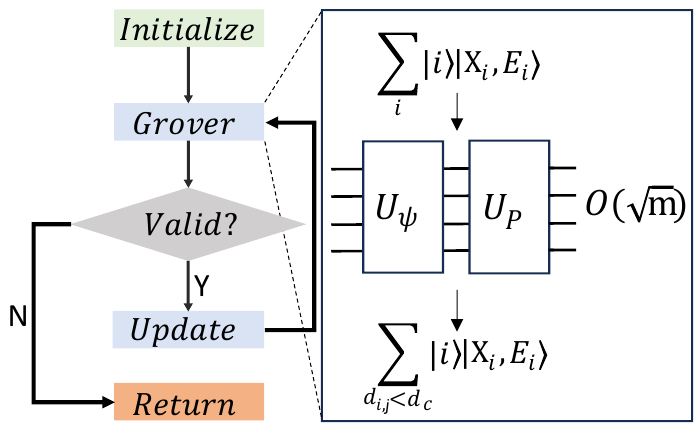}
    \caption{
        Algorithm flow for Local Density computation and for Assigning Clusters. The quantum state is initialized in the green ``Initialize" box. For Local Density Computation (Cluster Assignment), it comprises all points in the DSS $\mathcal{S}$ (in the DSS). The ``Grover" (light blue) block performs $U_\psi$ and $U_P$ in succession $O(\sqrt{m})$ times, and returns all points satisfying the required condition. The inset considers the case of Local Density computation where the condition is $d_{i,j}<d_{\rm c}$. For the cluster assignment step, we check if points in the DSS are followers of the points in the cluster $\mathcal{C}$ (see Sec.~\ref{subsec: cluster}). The output of the Grover subroutine is then measured to yield an index that is checked for validity in the grey ``Valid?" diamond.
        %Here, the index is valid if it satisfies the condition discussed above and in the main text. 
        If the point satisfies the chosen condition, the $Y$ branch is executed. Within the ``Update'' (light blue) step this point is then removed from either $\mathcal{S}$ or the DSS and stored to be returned in the ``Return" orange box. Once all points are found, the ``Valid?" condition triggers the $N$ branch to terminate the algorithm. Depending on the chosen subroutine, the returned indices are employed to compute the Local Density from Eq.~\eqref{eq: LE_computation}, or to construct $\mathcal{C}$.
    }
\label{fig:LD_flow}
\end{figure}

Once seeds and outliers are determined, the clusters are constructed. From the seeds, we iteratively combine ``followers". If point $\mathcal{N}_j$ is the Nearest Higher of point $j$, then point $j$ is termed as $\mathcal{N}_j$'s follower. The follower of a point is most likely generated by the same process as the point itself (in the context of ER, by the same particle), and as such shall be included in the same cluster. In Fig.~\ref{fig:full_algo}(c), the orange and purple points form two different clusters, and the followers of the points in the purple one are indicated by arrows.
% \end{enumerate}

\subsection{Local Density Computation}
\label{subsec: local}
% \begin{enumerate}
%     \item Local Density HEP intuition
%     \item Explanation of variables and mathematical definition with ref to figure 3
%     \item Flowchart (in the figure, with many references in the next point)
%     \item Explanation of flowchart steps
%     \begin{enumerate}
%         \item initialization
%         \item Grover
%         \item Measurement
%     \end{enumerate}
%     \item Brief description of Grover BlackBox (?) (say details can be found in sec 2.5) - I would say that in Sec. 2.5 you give a thorough description, and here you discuss about specific settings... Can include a circuit next to the flowchart to help
    
% \end{enumerate}

% As mentioned above, the data available at the end of collider events is the energies deposited by particles on the detectors (and the detector locations). When a particle travels through a region, it deposits higher energy on detectors near it but it is also possible for random detectors to be triggered during an event. Therefore, it makes sense to define a measure that combines the energies of the detectors around a point with that of the detector at the point. \\

In this section, we describe the subroutine (schematically represented in Fig.~\ref{fig:LD_flow}) that computes the Local Density $\rho_j$ of the point $j$, as defined in Eq.~\eqref{eq: LE_computation}. To perform the computation, all points $i$ whose distance $d_{i,j}$ from point $j$ is smaller than the threshold $d_{\rm c}$ need to be determined from the search space $\mathcal{S}$. This search space is the smallest set of tiles $\square_k$ required to cover the $d_{i,j}< d_{\rm c}$ consideration circle. In  Fig.~\ref{fig:full_algo}(a), $\mathcal{S}$ is highlighted in light blue.
%  Fig.~\ref{fig:full_algo}(a) pictorially describes the region under consideration for the Local Density computation. The base point indexed by $b$ is black and is at the center of the blue circle. This circle has a radius $d_{\rm c}$ and represents the region where $d_{i,I}<d_{\rm c}$. The points in this region are green while the rest in the search space are grey. The search space spans the two tiles shown in light blue.

We shall refer to $\mathcal{S}$ as the local dataset that, as explained above, can be efficiently prepared with the qRAM \cite{Giovannetti_2008}. To do so, we only require determining the tiles $\square_k$ that are in the search space, which can be done efficiently classically \cite{clue}. The initial state of this subroutine, after being prepared via the qRAM, is therefore
\begin{equation}
    \label{eq:ld_init}
    \sum_{k \in \mathcal{S}} \sum_{i \in \square_k} \ket{i} \xrightarrow{\text{qRAM}} \sum_{k \in \mathcal{S}} \sum_{i \in \square_k} \ket{i} \ket{X_i,E_i},
\end{equation} 
where the index $i$ is unique for each point in $\mathcal{S}$. $i \in \square_k$ indicate all indices within tile $k$ [either of the light blue squares in Fig.~\ref{fig:full_algo}(a)]. Ancillary qubits, omitted for clarity in Eq.~\eqref{eq:ld_init}, are employed within the Grover search (for more information, see  App.~\ref{sec: Grover}).
%
% The quantum state from Eq.~\eqref{eq: dataset} 
%
%
% where $(x_i,y_i)$ are the coordinates of the detector $i$, $E_i$ is the energy reading and $A_i$ are ancilla qubits for the computation and the summation is carried out over all detectors in the search space (in light blue in  Fig.~\ref{fig:full_algo}(a). The encoding is FLOAT for all registers except $\ket{i}$.

At this stage, we must find the points $i$ [green dots in Fig.~\ref{fig:full_algo}(a)] that are within a radius of $d_{\rm c}$ from the base point $j$ [black point in Fig.~\ref{fig:full_algo}(a)].
As shown in  Fig.~\ref{fig:LD_flow}, we perform Grover Search to prepare \cite{Brassard_2002}
%, with high probability \cite{Brassard_2002},
%
\begin{equation}
\label{eq: good_indices_ld}
     \sum_{i} \ket{i} \ket{X_i, E_i} \xrightarrow[]{\rm Grover} \sum_{d_{i,j} < d_{\rm c}} \ket{i} \ket{X_i, E_i}.
\end{equation}
Here, the first register of the Grover output contains all points characterized by indices $i$ such that $d_{i,j}<d_{\rm c}$. As shown in the inset of the figure, the Grover Search consists of $O(\sqrt{m})$ repetitions (where $m$ is the number of points in $\mathcal{S}$) of the $U_{\psi}$ and $U_P$ operators. $U_P$ is the diffusion operator and $U_{\psi}$ is the unitary associated with the oracle of Grover Search \cite{grover}. Further details regarding Grover Search and the unitaries we use for our algorithm can be found in App.~\ref{sec: Grover}. 

When the algorithm is run, measurement either yields a point that satisfies this distance condition, or (if there are no valid indices left) an index that does not satisfy this condition. This is verified by the grey ``Valid?" diamond in Fig.~\ref{fig:LD_flow}. The branched logic following this block ensures that the algorithm loops until all the required points are returned by the algorithm in the ``Return" block. 

Once we have obtained all indices $i$ of points satisfying the distance condition ($d_{i,j} < d_{\rm c}$), we perform the summation in Eq.~\eqref{eq: LE_computation}. This is computed and stored in the original dataset for each point. The database is now updated using qRAM with local density values for all points where the $j^{\rm th}$ point in the database has the corresponding computed local density $\rho_j$.

The scaling of the subroutine that determines the local density of a single point is given by the number of points in the blue consideration circle in Fig.~\ref{fig:full_algo}(a) such that $d_{i,j}<d_{\rm c}$. If we say this number is $p$, $O(p)$ runs are required. This is therefore a $O(p \sqrt{m})$ algorithm as opposed to the $O(m)$ classical iterative algorithm for the Unstructured Search Problem. 
% \footnote{We can either ensure that the black box does not mark found points, or that the qRAM excludes that point from the input of the next Grover iteration.}

As a final remark, we highlight that it is in principle possible to design a unitary that computes the Local Density directly and stores the output in a quantum register. This unitary would remove the requirement of finding individually the indices $i$ such that $d_{i,j} < d_{\rm c}$, thus removing the overhead of $p$ in $O(p \sqrt{m})$. However, designing this circuit is non-trivial and its depth may be large. This is therefore left for future investigations.

\subsection{Find Nearest Higher}
\label{subsec: nh}
%
% Since Reconstruction is a clustering task, it is helpful to think of a procedure that allows assigning a point to a cluster or to the set of outliers in case it is unlikely to belong to any cluster. The highest local density point in a cluster is most likely to be the position through which the particle passed and is termed a Seed. It also makes sense to assign special significance to the nearest point with a higher Local Density as traversal from point to corresponding seed would be possible through these special points. These points are termed Nearest Highers. \\
%
Here, we describe qLUE's subroutine for finding the Nearest Highers ($\mathcal{N}_{j}$) introduced in Sec.~\ref{subsec: overview}. As a reminder, $\mathcal{N}_j$ is the nearest point to the base point $j$ whose local density $\rho_{\mathcal{N}_j}$ is more than the local density $\rho_j$ of the base point, see Eq.~\eqref{eq:seed}.
\begin{figure}[!htb]
    % \centering
    \includegraphics[width=\columnwidth]{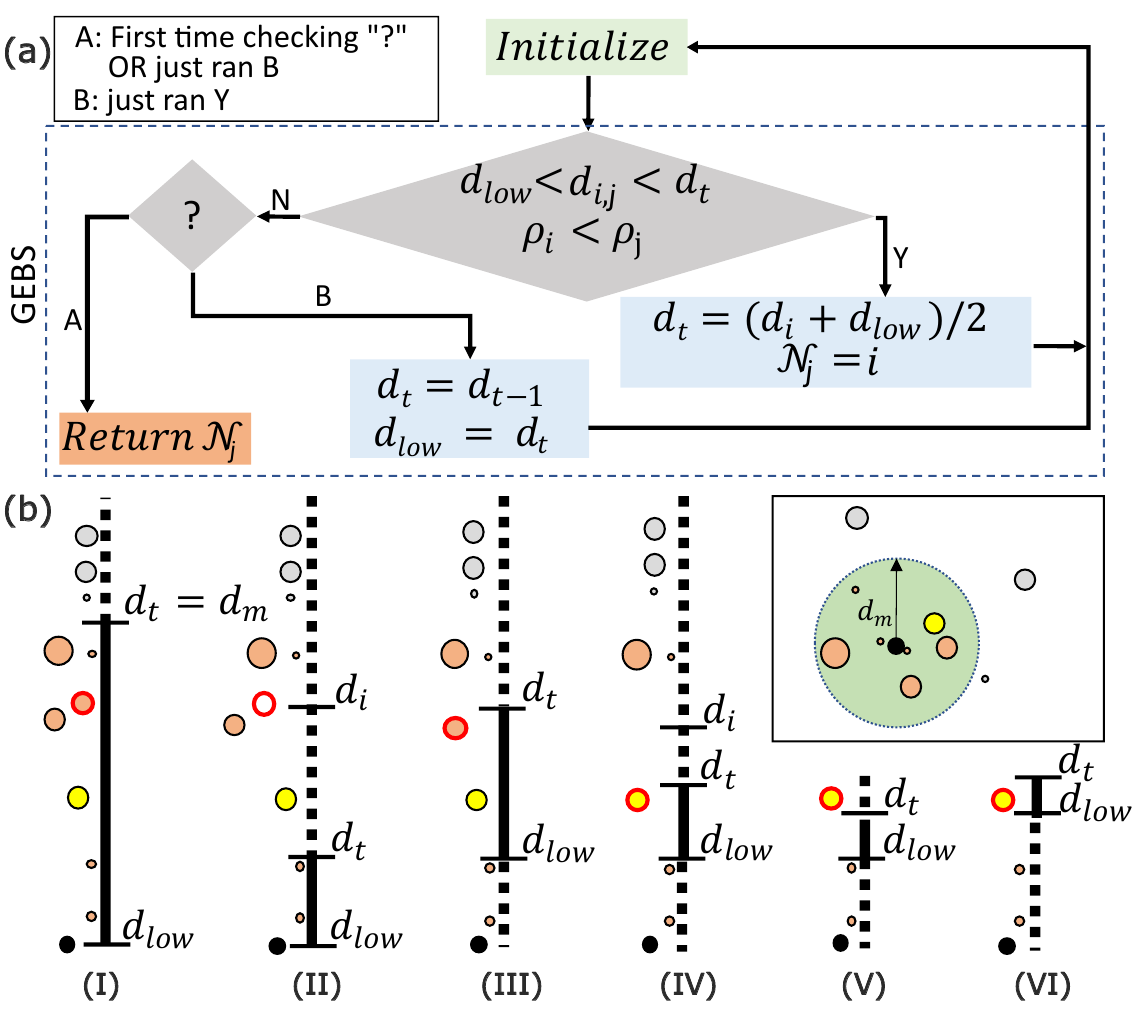}
    \caption{
    (a) Diagrammatic representation of the algorithm. GEBS determines successive candidates for the ``Nearest Higher'' until the proper one is found. The quantum state in Eq.~\eqref{eq: nh_init} is prepared in the ``Initialize'' step (green box). Grover Search (larger diamond) is then performed to find the points satisfying $d_{\rm L} < d_{i,j} < d_{\rm t}, \rho_i < \rho_j$. If this condition is satisfied (`$Y$' branch), $d_{\rm t}$ is updated and Grover run again. If not (`$N$' branch), control flows to the ``?" diamond. The branch $A$ is entered if the ``?" condition is being checked for the first time or if branch $B$ was just run. Branch $B$ is entered if branch $Y$ was just run. (b) The algorithm's working is shown step-by-step (numbers at the bottom) for the search space $\mathcal{S}$ in the inset in the top right corner. The points are mapped to a line where the height represents the distance $d_{i,j}$ from the base point $j$ (black dot at the bottom). The grey (orange) points are outside (inside) the green consideration circle with radius $d_{\rm m}$ [see also Fig.~\ref{fig:full_algo}(b)]. At each step of GEBS, the thresholds $d_{\rm L}$ and $d_{\rm t}$ are updated according to the logic in panel (a). The dot with the red border indicates the current candidate for $\mathcal{N}_j$; when filled (empty) it is (not) found by Grover Search at that step. The yellow point is the Nearest Higher $\mathcal{N_j}$ that is found at the end of GEBS.
    }
    \label{fig:nh_flow}
\end{figure}

Similar to the initialization carried out for the Local Density Computation step, we use qRAM to initialize the quantum state
\begin{equation}
\label{eq: nh_init}
    \sum_{k \in \mathcal{S}} \sum_{i \in \square_k} \ket{i} \xrightarrow{\text{qRAM}} \sum_{k \in \mathcal{S}} \sum_{i \in \square_k} \ket{i} \ket{X_i} \ket{\rho_i}.
\end{equation}
Here, the indices $i$ are within the tiles $\square_k$, as in Eq.~\eqref{eq:ld_init}, and $\mathcal{S}$ is the considered search space, schematically represented by the light green box in Fig.~\ref{fig:full_algo}(b). This search space is determined from $d_{\rm m}$ as opposed to $d_{\rm c}$, which is the user-defined threshold that is set to be $\delta d_{\rm c}$. Note that the energy $E_i$, employed for determining the densities $\rho_i$ in Sec.~\ref{subsec: local}, is hereon not required.

% The ancilla set $A_i$ consists of registers for $x_b, y_b, \Tilde{\rho}$ as well as registers to store intermediate computation results and the final answer as explained in App.~\ref{subsec: bbop} and in Fig.~\ref{fig:blackbox_operator}.

To find the Nearest Higher, we use a Grover-Enhanced Binary Search (GEBS) where each search step is enhanced by Grover's algorithm. The output of every Grover run,
\begin{equation}
    \sum_{\substack{d_{\rm L}<d_{i,j}<d_{\rm t}, \\ \rho_i > \rho_b}} \ket{i} \ket{{X_i, \rho_i}},
    \label{eq: good_NH}
\end{equation}
is a superposition over all points $i$ whose distance $d_{i,j}$ from the base point $j$ lies between the thresholds $d_{{\rm L}}$ and $d_{\rm t}$. Furthermore, their local density $\rho_i$ should be higher than that of the base $\rho_j$. At each step, $d_{\rm L}$ and $d_{\rm t}$ are updated based on whether a point satisfying the conditions in the grey diamond of Fig.~\ref{fig:nh_flow}(a) is found. Ancilla registers are used here as detailed in App.~\ref{sec: Grover}.

To better understand the algorithm, we provide a step-by-step walkthrough of the example in Fig.~\ref{fig:nh_flow}(b). The search space $\mathcal{S}$ 
%state in Eq.~\eqref{eq: nh_init} 
is schematically represented by the inset in the right hand side, where each dot represents a point with a size that is proportional to its local density. The consideration circle (light green, dotted border) highlights all points within a radius $d_{\rm m} = \delta d_{\rm c}$. In this work, we set the outlier delta factor $\delta$ to $2$. The consideration circle in the inset corresponds to $d_L=0$ and $d_{\rm t}=d_{\rm m}$, shown in step (I). In the main panel, vertical lines refers to the steps (I-VI) of GEBS that are reported below, and schematically represent the distances of all points (coloured dots) from the base point $j$ (black one at the bottom).

GEBS starts with the higher threshold set as $d_{\rm t} = d_{\rm m}$ and the lower threshold $d_{\rm L} = 0$ as shown in vertical line (I) of Fig.~\ref{fig:nh_flow}(b). Following the probabilistic nature of quantum mechanics, assume that the point with a red border indexed $i$ is found after measuring the output of the Grover Search in Eq.~\eqref{eq: good_NH}. This triggers the updates in the $Y$ branch in the diagram of Fig.~\ref{fig:nh_flow}(a), such that we assign $\mathcal{N}_j = i$ and update $d_{\rm t} \mapsto (d_{i,j} + d_{\rm L})/2$ . The point indexed $i$ is then removed from the search space, as can be seen in (II).
Now, since no point satisfies the conditions in the diamond of the flow diagram [see (II)] and $d_{\rm t}$ was just set to $(d_{i,j} + d_{\rm L})/2$, the $B$ branch is carried out. This updates the thresholds $d_{\rm t}$ and $d_{\rm L}$ for the next iteration of the algorithm, see (III).

Now, assume that the new point with a red border is found [step (III)]. Updates in the $Y$ branch of Fig.~\ref{fig:nh_flow}(a) are carried out again with a new index $i$ and the search region is reduced to contain a single point. In the next step (IV), that point (yellow) is found and, for the third and last time, the nearest higher and the thresholds are triggered according to the $Y$ branch. Next, since no point is found in (V), qLUE executes the updates in the $B$ branch of the diagram.
In the last iteration (VI), no points satisfy the desired conditions. The parameter $d_{\rm t}$ was just set to $d_{{\rm t}-1}$, i.e, the subroutine just ran $B$ which means that the $A$ branch is now executed and $\mathcal{N}_j$ is returned. 
%Currently, this is the yellow point with a red outline in the last vertical line of Fig.~\ref{fig:nh_flow}(b). This point is therefore returned as the Nearest Higher.

% Consider that each step of GEBS consists of searching among a set of at most $m$ points and there are $n$ points in the search space.
The runtime complexity of the GEBS procedure, with $m$ points in the search space $\mathcal{S}$, is $O(\alpha\sqrt{m})$ as opposed to $O(m)$ classically. The $\alpha$ term is due to the binary search procedure and depends on the size of the quantum register used to encode the distance. Specifically, for a chosen precision $2^{-\Delta}$ used for the positions of the points in the datasets, $\alpha = \Delta$.

% \begin{figure}[!htb]
%     \centering
%     \includegraphics[scale=0.4]{groverhalfing.png}
%     \caption{This figure shows the step-by-step working of the Grover-Enhanced Binary Search algorithm.}
%     \label{fig:gh}
% \end{figure}

%
\subsection{Find Seeds, Outliers, and Assign Clusters}
\label{subsec: cluster}
Once the Nearest Highers $\mathcal{N}_j$ are determined for all points $j$ in the dataset, Seeds and Outliers %(see Sec.~\ref{subsec: overview} for their definitions) 
are found via another Grover Search over all points in the dataset. As per the definition in Eq.~\eqref{eq:seed}, Seeds [red points in Fig.~\ref{fig:full_algo}(c)] are the points with highest local density within a neighbourhood. Outliers [blue points in Fig.~\ref{fig:full_algo}(c)] are mathematically described by Eq.~\eqref{eq:outlier}, are most likely noise, and therefore do not belong to any cluster.

Similar to the previous subroutines, the quantum registers for these procedures are initialized via qRAM. Seeds and outliers are then determined based on the corresponding conditions via Grover Search. Two quantum registers, the first marking whether a point is an outlier and the second to store the seed number -- which is also the cluster number -- are added to the quantum database.

The final subroutine of qLUE is the assignment of points to clusters. At this stage, outliers have been removed from the input dataset, as they have been already identified. The algorithm flow is the same as that of the Local Density step in Fig.~\ref{fig:LD_flow}.
For a chosen seed $s$, we define $\mathcal{C}$ to be the set containing the indices of all points determined to be in the associated cluster at the end of this subroutine. To assign points to $\mathcal{C}$, we follow a procedure similar to that of the Local Density step in Fig.~\ref{fig:LD_flow}. In the ``Initialize" step, $\mathcal{C}$ is initialized to $\{s\}$ and the quantum registers are initialized via qRAM to the state
\begin{subequations} 
\label{eq: seed_outlier}
\begin{align}
    & \sum_{i \in \text{DSS}} \ket{i} \xrightarrow{\rm qRAM} \sum_{i} \ket{i} \ket{{\rm V}_i}, \label{eq:seed_outlier_a}\\ 
    & \ket{{\rm V}_i} = \ket{X_i, \rho_i, d_{\mathcal{N}_i,i}, X_{\mathcal{N}_i}}.
\end{align}
\end{subequations}

In the ``Grover" block, we search over a superposition of points in the dataset which we call the Dynamic Search Space (DSS). The DSS differs from the search space $\mathcal{S}$ in the Local Density step as it is dynamic. This is because it depends on the points in $\mathcal{C}$, which are updated at each iteration. In Fig.~\ref{fig:full_algo}(c), for instance, the red seed and the orange point both with black borders are the elements of the current $\mathcal{C}$. To find the DSS, a square window of edge $2d_{\rm m}$ is first opened for every point in $\mathcal{C}$ (in the figure, the squares with the same border style as the corresponding points). A rectangular region (red box) is then obtained by finding the axis-aligned minimum bounding box for these windows. The set of tiles $\square_k$ covered partially or fully by this minimum bounding box is the DSS. For example, in Fig.~\ref{fig:full_algo}(c), it comprises the $9$ tiles highlighted in light red.

With a similar procedure as for the Local Density subroutine, the ``Grover" block now systematically identifies all followers of all points within set $\mathcal{C}$. Here, in the ``Update" step in Fig.~\ref{fig:LD_flow}, as the point found by the ``Grover" block has passed the ``Valid" condition, it is appended to $\mathcal{C}$. Once no more points are found, the ``Return" block yields $\mathcal{C}$, following the same flow as the Local Density computation subroutine.

% If point $i$ has nearest higher $d'_i$, it is termed a \textit{follower} of $NH_i$.
% Grover's Algorithm is now used to search for the $m$ followers of the points in $\mathcal{C}$ among the $n$ points in the opened window. This step provides a quadratic speedup as it has a complexity of $O(m\log m \sqrt{n})$ as opposed to the classical $O(n)$ complexity of search in each window.

% In the APPEND step, the follower indices are appended to $\mathcal{C}$. Program flow then goes back to the UPDATE DS step. The algorithm stops when Grover no longer finds any followers. Note that as $\mathcal{C}$ grows, so does the size of the window.

The complexity of the Cluster Assignment step is 
%dependent on the dataset, distribution of the points in it and their corresponding energies, we focus only quantifying the advantage the quantum algorithm provides here which is 
similar to the one of the Local Density Computation subroutine. The quantum advantage stems from the quadratic speedup provided by the Grover algorithm, which allows determining the follower faster if compared to CLUE. If there are $f$ points in a cluster $\mathcal{C}$ and $m$ points in the corresponding DSS, the classical complexity of the Cluster Assignment step is $O(m)$, while the quantum algorithm has a runtime of $O(f\sqrt{m})$.
%similar to the advantage Grover Search provides in the Local Density computation step. Concretely, this advantage appears in finding the followers of the points in cluster $\mathcal{C}$. If there are $f$ followers and $m$ points in the DSS corresponding to $\mathcal{C}$, the classical complexity is $O(m)$ while the quantum algorithm has a runtime of $O(f\sqrt{m})$.

%
\section{Results}
\label{sec:results}

In this section, we test qLUE in multiple scenarios, each designed to investigate its performance for different settings.
In Sec.~\ref{subsec: scoring}, we introduce the scoring metrics used for our analysis. In Sec.~\ref{subsec: noise}, we describe the performance of the algorithm applied on a single cluster in a uniform noisy environment. In Sec.~\ref{subsec: overlap}, we study the performance on overlapping clusters.
%with the number of points in one cluster ($N_1$) being an integer multiple of the number of points in the other ($N_2$), i.e, $N_1 / N_2 \in \mathbb{Z}$ 
Finally, in Sec.~\ref{subsec: non_centroidal}, we study the performance of qLUE on non-centroidal clusters with and without an energy profile.

\begin{figure*}[!htb]
    \centering
    \includegraphics[width=\textwidth]{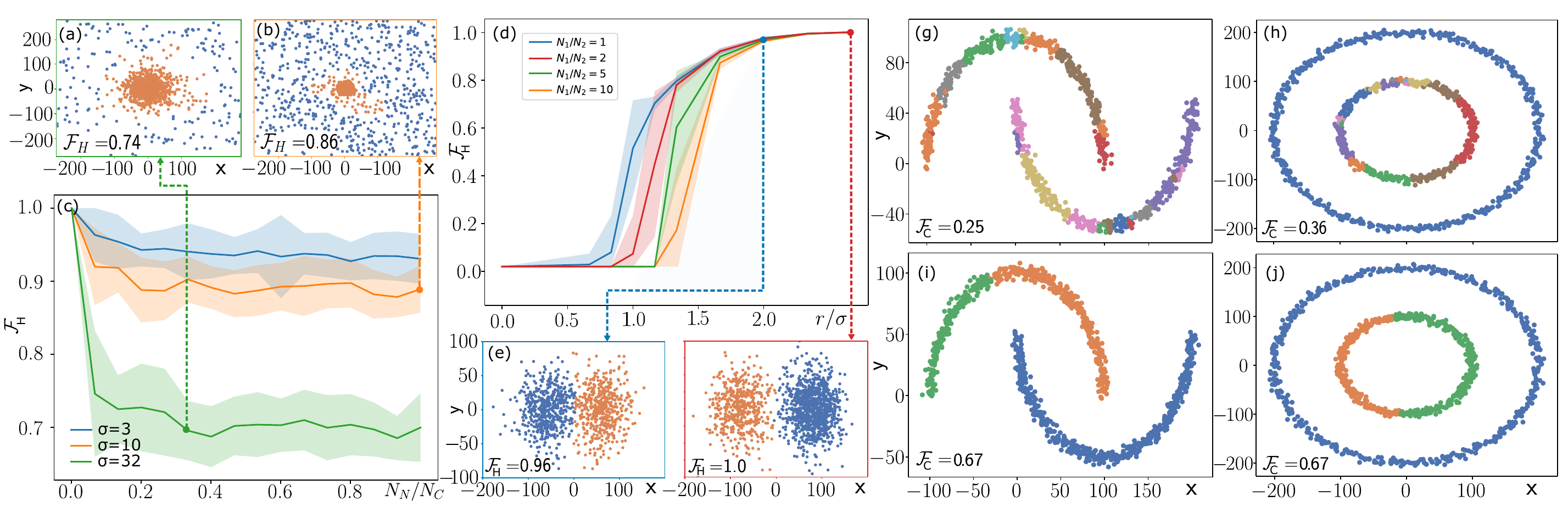}
    \caption{Numerical results from qLUE simulated on a classical machine.
    %that study the effect of noise and overlap on clustering performance as well as show the effect of an appropriately chosen energy profile in clustering the non-centroidal moons and circles datasets. 
    (a-c) qLUE's performance in noisy environments. The dataset generated for these experiments and visualized in panels (a) and (b) consists of a cluster (noise) with $N_C = 750$ ($N_N$) points sampled from the Gaussian distribution in Eq.~\eqref{eq: gauss} 
    %with covariance matrix  
    %$\Sigma = \begin{bmatrix}
    %    \sigma^2 & 0 \\
    %    0 & \sigma^2 
    %\end{bmatrix}$ 
    %and centered at the origin 
    (uniform distribution) over a square of size $500$. The energy of noise points is sampled uniformly between zero and one, while each cluster point is assigned an energy that is the probability of being sampled multiplied by a factor $A = 500$. (a-b) Computed clusters at $N_N/N_C = 0.33, \sigma=32$, and $N_N/N_C = 1, \sigma=10$, respectively. In (c), $\mathcal{F}_H$ is plotted against $N_N/N_C$ for the $\sigma$ in the legend. 
    (d-f) Performance for overlapping clusters. In (d), $\mathcal{F}_H$ vs $r/\sigma$ is shown for $\sigma = 30$ and different ratios $N_1/N_2$. Here, $r$ is the distance between the centers of two clusters with $N_1 = 500$ and $N_2$ points, and we assign to each point an energy that is equal to its sampling probability in Eq.~\eqref{eq: gauss}. (e-f) Computed clusters at $r/\sigma = 2.0, N_1/N_2=1$, and $r/\sigma = 2.67, N_1/N_2=2$, respectively. 
    The shadowed regions in (c-d) represent the standard deviations of $\mathcal{F}_H$ over $30$ iterations. 
    (g-j) Performance over non-centroidal clusters of $500$ points each generated from $scikit-learn$ \cite{pedregosa2018scikitlearn}. In (g-h) the points' energy profile is uniform, while in (i-j) is varied linearly with respect to the distance such that each cluster has a single, most energetic point (see Sec.~\ref{subsec: non_centroidal}). For all experiments, $d_{\rm c}$ was set to $20$ and $\Tilde{\rho}$ was set to $25$.
    (a-f) use the energy-aware metric in Eqs.~\eqref{eq: homogeneity} while in (g-j), since the energy profile is assigned by the user and is not part of the dataset itself, in the scoring process we set all points to have the same energy.
    }
    \label{fig:result}
 \end{figure*}
    % TYh\caption{From L-R, top then bottom: i. Gaussian clusters separated by a distance, ii. Homogeneity score vs Distance between Gaussian centers plot for various energy factors between the two clusters, iii. Gaussian cluster with noise, iv. Homogeneity vs number of noise samples (300 cluster samples throughout the experiment), v. Results of qLUE on data with no variation in energy, vi. Results of qLUE on data with energy factors \{1,10,100\}, vii. Results of qLUE on the Moons dataset, viii. Results of qLUE on vi but with added noise, ix. Results of qLUE on nested circles data, x. Results of qLUE in the overlapping noisy setting with the same energy distribution as vi and viii }
    % 

\subsection{Scoring metrics: Homogeneity and Completeness scores}
\label{subsec: scoring}
To weigh more energetic points such as seeds higher than the others, we use modified, energy-aware versions \cite{Jaroslavceva:2865866} of the Homogeneity ($\mathcal{F}_H$) and Completeness ($\mathcal{F}_C$) scores \cite{rosenberg2007v}. These metrics are defined in terms of the predicted cluster labels $\mathcal{C}_{\rm p}$ obtained from qLUE, and the true cluster labels $\mathcal{C}_{\rm t}$ of the generated dataset. $\mathcal{F}_H$ and $\mathcal{F}_C$ are based on the energy aware \cite{Jaroslavceva:2865866} mutual information $I(\mathcal{C}_{\rm p}:\mathcal{C}_{\rm t})$, the Shannon entropy $H(\mathcal{C}_{\rm t})$, and the joint Shannon entropy $H(\mathcal{C}_{\rm t}, \mathcal{C}_{\rm p})$ \cite{Nielsen-Chuang-2010}:
%
% \begin{subequations} 
% \label{eq: homogeneity}
% % \centering
% \begin{align}
%     & \mathcal{F}_H = \frac{I\left(\mathcal{C}_{\rm p}:\mathcal{C}_{\rm t}\right)}{H\left(\mathcal{C}_{\rm t}\right)} 
%     \text{ and }
%     \mathcal{F}_C = \frac{I\left(\mathcal{C}_{\rm p}:\mathcal{C}_{\rm t}\right)}{H\left(\mathcal{C}_{\rm p}\right)}
%     , \\
%     & H\left(\mathcal{C}_{\rm p}\right) = -\sum_{a} P\left(\mathcal{C}_{\rm p} \overset{?}{=} a\right) \log_2 P\left(\mathcal{C}_{\rm p} \overset{?}{=} a\right)
%     , \\
%     & H\left(\mathcal{C}_{\rm t}\right) = -\sum_{b} P\left(\mathcal{C}_{\rm t} \overset{?}{=} b\right) \log_2 P\left(\mathcal{C}_{\rm t} \overset{?}{=} b\right)
%     , \\
%     & H\left(\mathcal{C}_{\rm p}, \mathcal{C}_{\rm t}\right) = -\sum_{a}\sum_{b} P\left(a, b\right) \log_2 P\left(a, b\right)
%     , \\
%     & I\left(\mathcal{C}_{\rm p}:\mathcal{C}_{\rm t}\right) = H\left(\mathcal{C}_{\rm p}\right) + H\left(\mathcal{C}_{\rm t}\right) - H\left(\mathcal{C}_{\rm p}, \mathcal{C}_{\rm t}\right)
%     .
% \end{align}
% \end{subequations}

\begin{subequations} 
\label{eq: homogeneity}
% \centering
\begin{align}
    & \mathcal{F}_H = \frac{I\left(\mathcal{C}_{\rm p}:\mathcal{C}_{\rm t}\right)}{H\left(\mathcal{C}_{\rm t}\right)} 
    \text{ and }
    \mathcal{F}_C = \frac{I\left(\mathcal{C}_{\rm p}:\mathcal{C}_{\rm t}\right)}{H\left(\mathcal{C}_{\rm p}\right)}
    , \\
    & H\left(\mathcal{C}_{\rm p}\right) = -\sum_{a} \frac{E_a}{E} \log_2 \frac{E_a}{E}
    , \\
    & H\left(\mathcal{C}_{\rm t}\right) = -\sum_{b} \frac{E_b}{E} \log_2 \frac{E_b}{E}
    , \\
    & H\left(\mathcal{C}_{\rm p}, \mathcal{C}_{\rm t}\right) = -\sum_{a}\sum_{b} \frac{E_{a,b}}{E} \log_2 \frac{E_{a,b}}{E}
    , \\
    & I\left(\mathcal{C}_{\rm p}:\mathcal{C}_{\rm t}\right) = H\left(\mathcal{C}_{\rm p}\right) + H\left(\mathcal{C}_{\rm t}\right) - H\left(\mathcal{C}_{\rm p}, \mathcal{C}_{\rm t}\right)
    .
\end{align}
\end{subequations}
As discussed in \cite{Jaroslavceva:2865866}, $E_a$ is the energy aggregated over all points that qLUE classifies into cluster $a$. $E_b$ is the energy aggregated over all points in cluster $b$ in the true dataset. $E_{a,b}$ is the sum of energies of all points in cluster $b$ in the true dataset that are also assigned to cluster $a$ by qLUE. $E$ is the accumulated energy of all points in the dataset. We remark that for unit energies, Eqs.~\eqref{eq: homogeneity} reduce to the more common form presented in Ref.~\cite{rosenberg2007v}.
%The above definitions are adapted from \cite{Jaroslavceva:2865866}.

qLUE applied to an input dataset yields homogeneity $\mathcal{F}_H = 1$ if all of the predicted clusters only contain data points that are members of a single true cluster. On the other hand, $\mathcal{F}_C = 1$ is obtained if all the data points that are members of a given true cluster are elements of the same reconstructed cluster. Therefore, these metrics are better suited to different scenarios. The impacts of noise and cluster overlap investigated in Secs.~\ref{subsec: noise} and \ref{subsec: overlap} are better captured by $\mathcal{F}_H$. Indeed, if qLUE incorrectly classifies noise points into predicted clusters, $\mathcal{F}_C$ is unaffected. 
%but $\mathcal{F}_H$ will be lowered. 
On the other hand, $\mathcal{F}_C$ shall be employed when studying non-centroidal clusters in Sec.~\ref{subsec: non_centroidal}, since $\mathcal{F}_H = 1$ if one true cluster is divided by qLUE into many sub-clusters.

% So, as we will see in the following sections it makes sense to use $\mathcal{F}_H$ for Fig.~\ref{fig:result}(a-f) and $\mathcal{F}_C$ for Fig.~\ref{fig:result}(g-j).

%
\subsection{Noise}
\label{subsec: noise}
Here, we study the performance of qLUE for a single cluster in a noisy environment. We vary the number $N_N$ of noise points sampled from a uniform distribution over a square region of fixed size. A cluster of $N_C$ points with coordinates $X_j = \left[ x_{j,1}, x_{j,2} \right]$ is sampled from the multivariate Gaussian distribution
%whose probably density function $pdf(X)$ is
%
\begin{equation}
\label{eq: gauss}
pdf(X_j) = \frac{e^{-\frac{1}{2}(X_j-\mu_j)^T \Sigma^{-1} (X_j-\mu_j)}}{(2\pi)^{\frac{n}{2}}|\Sigma|^{\frac{1}{2}}} ,
\end{equation}
where $\mu =\left[ \mu_{x_1} , \mu_{x_2} \right]^T$ is the mean of the distribution (set to $[0,0]^T$ in our case) and $\Sigma$ the covariance matrix. Here, we choose $\Sigma = \sigma \mathcal{I}$, with $\mathcal{I}$ being the identity matrix and $\sigma$ a positive real number.
%, and $\mu$ is the centre of the considered square.

Examples of the generated clusters (in orange) and noise (in blue) are given in Fig.~\ref{fig:result}(a-b) for $N_N/N_C = 0.33$ at $\sigma = 32$ and $N_N/N_C = 1$ at $\sigma = 10$, respectively. The energy assigned to each point $X_j$ in the cluster is given by $A \times pdf(X_j)$ [see Eq.~\eqref{eq: gauss}] with $A = 5 \times 10^2$. The energy of each noise point is randomly sampled between zero and one. This choice resembles the typical scenarios in ER tasks which CLUE \cite{clue} was designed for.

In Fig.~\ref{fig:result}(c), we show the variation of homogeneity score $\mathcal{F}_H$ with respect to the ratio $N_N/N_C$. We employ the values of $\sigma$ reported in the legend, that are associated to different colours in the plot. As can be seen, the clustering performance is inversely proportional to both $N_N/N_C$ and $\sigma$. When these parameters are small, the typical distance between cluster points is much smaller than that between noise points, and $\mathcal{F}_H$ approaches unity. With a higher chance of labeling noisy points as within the cluster, however, $\mathcal{F}_H$ is lowered. As such, the degrading of $\mathcal{F}_H$ is proportional to the probability of a noise point being in the cluster region, which increases with both $\sigma$ and $N_N/N_C$.

\subsection{Overlap}
\label{subsec: overlap}
Here, we consider the case of two circular clusters with $N_1$ and $N_2$ points respectively, each sampled from the multivariate Gaussian distribution in Eq.~\eqref{eq: gauss} and with $\Sigma = \sigma \mathcal{I}$. The energy profile is determined by $ pdf(X_j)$ for coordinates $X_j$. 
%and with the same value of $50000$ for $A$. 
The centers $\mu_1$ and $\mu_2$ (two instances of $\mu$) are chosen to be $(r/2,0)$ and $(-r/2,0)$, respectively, such that the distance between the cluster centers is $r$.

In Fig.~\ref{fig:result}(d), we study the variation of homogeneity score $\mathcal{F}_H$ as a function of $r/\sigma$ for several values of $N_2/N_1$. The computed clusters for $r/\sigma = 2$ at $N_2/N_1 = 1$ and $r/\sigma = 2.67$ at $N_2/N_1 = 2$ are shown in panels (e) and (f), respectively, to showcase the typical scenarios considered here. 

For all $N_1/N_2$, $\mathcal{F}_H$ is zero for low $r/\sigma$ (high overlap). There is then a region where $\mathcal{F}_H$ increases with $r/{\sigma}$ and then saturates at unity for high $r/\sigma$ (little to no overlap). When the two clusters are too nearby, i.e., $r/\sigma \ll 1$, they are in fact indistinguishable and qLUE labels all points together. Increasing the ratio $r/\sigma$ makes the clusters move away from each other and thus qLUE can discern them. Importantly, large values of $\mathcal{F}_H$ are already attained when the clusters still have a significant overlap. In this scenario, employing the energy labels and the energy density considerably contributes to assigning the points to the right cluster. In fact, the nearest higher points are more likely to connect the points near or on the decision boundary with the more energetic core, thus separating the clusters better. 

The performance of qLUE is also affected by the ratio $N_1/N_2$. When one cluster contains more points than the other, it is more likely to ``capture" points from the smaller. The resulting loss in homogeneity score $\mathcal{F}_H$ for low $r/\sigma$ ratios is evident from Fig.~\ref{fig:result}(d), where it can be seen that clusters of similar sizes are better distinguished from each other.

\subsection{Non-centroidal Clusters}
\label{subsec: non_centroidal}
Finally, we study the performance of qLUE on non-centroidal clusters. For this purpose, we use the Moons and Circles datasets in Fig.~\ref{fig:result}(g-j), generated using $scikit-learn$ \cite{pedregosa2018scikitlearn}. Two settings are considered - one where a uniform energy profile is applied over the points [panels (g-h)] and one where a linear gradient energy profile is employed [panels (i-j)]. 

In the latter case, for each cluster we assign the highest value of the energy to a single point and lower the energies of all other points proportionally to their $x_2$ coordinate. In the case of the moon dataset, $E = x_2$ for the upper moon (so the top point of the upper moon has the maximum energy in the cluster) and $E = 60 - x_2$ for the lower moon (so the bottom point has the highest energy in the cluster). For the circles, $E = |x_2-200|/10$ for the inner circle  and $E = |x_2+100|/5$ for the outer one.

Since these datasets have no noise and are well separated, $\mathcal{F}_H$ is always one and we employ $\mathcal{F}_C$ to characterize the performance of qLUE. As in Fig.~\ref{fig:result}(g-h) the energy profile is uniform, and several points satisfy the seed condition. Therefore, qLUE groups each circle into several clusters, such that we obtain limited values for $\mathcal{F}_C$. On the contrary, cases with an energy profile assigned [Fig.~\ref{fig:result}(i-j)] results in less seeds that are better recognized by qLUE, and the completeness score $\mathcal{F}_C$ is considerably enhanced.

\section{Conclusion and Outlook}
\label{sec:conclusions}
We introduced qLUE, a novel quantum clustering algorithm designed to address the computational challenges associated with high-dimensional datasets. The significance of qLUE lies in its potential to efficiently cluster data leveraging quantum computing, mitigating the escalating computational complexity encountered by classical algorithms as dimensions increase. The algorithm's ability to navigate high-dimensional spaces is particularly promising when the density of points is very large, such that local searches become too demanding for classical computers. Therefore, qLUE will be beneficial in multiple scenarios, ranging from quantum-enhanced machine learning \cite{Zeguendry2023-ss, Haug_2023} to complex data analysis tasks \cite{qml}.  %These scenarios include, among others, ER \cite{clue} and motion planning in Robotics \cite{petrović2018motion}.

According to our numerical results, qLUE works well and its performance is significantly enhanced when an energy profile is assigned. Specifically, we study qLUE in noisy environments, on overlapping clusters, and on non-centroidal datasets that are commonly used to benchmark clustering algorithms \cite{tiwari2020kernel, Fujita2021-rh}. In scenarios that are typically encountered in ER tasks, qLUE correctly reconstructs the true clusters to a high level of accuracy. On the other hand, an energy profile can significantly boost qLUE performance in the case of non-centroidal clusters. Our numerical results, backed up by the well-studied CLUE and by the quadratic speedup stemming from Grover search, make qLUE a promising candidate for addressing high-dimensional clustering problems \cite{weiharrowthaler, kerenidis, Magano2023}.    %Following the scaling laws outlined in Sec.~\ref{sec:introduction}, we expect qLUE to be particularly advantageous, if compared to other classical approaches, in higher dimensional datasets.

As a first outlook, we identify the implementation of qLUE on NISQ hardware \cite{lukin, Bernien2017, Labuhn2016, Arute2019, Lanyon2011, Debnath2016, Córcoles2015}. This requires a comprehensive consideration of real device constraints. Aspects such as circuit optimization \cite{Nash_2020}, and the impact of noise will be critical and must be carefully addressed. Second, it is possible to improve the scaling of qLUE by devising a unitary that mitigates the need for repeating Grover's algorithm and thereby eliminating the factors of $p$, $\alpha$, and $f$  in the scaling of the subroutines outlined in Secs.~\ref{subsec: local}, \ref{subsec: nh} and \ref{subsec: cluster} respectively. Finally, it is worth investigating variations of qLUE that improve the quality of clustering in different scenarios. For instance, one can devise more sophisticated criteria for the Nearest Higher or Local Density computation steps. One can also improve performance by performing exhaustive hyperparameter searches or via hyperparameter optimization algorithms \cite{WU201926}.

\section*{Acknowledgements}
We thank the CERN Quantum Initiative, Fabio Fracas for creating the fertile ground for starting this project and Andrew J. Jena as well as Priyanka Mukhopadhyay for theoretical support.
WR acknowledges the Wolfgang Gentner Programme of the German Federal Ministry of Education and Research (grant no. 13E18CHA).
LD acknowledges the EPSRC quantum career development grant EP/W028301/1. DG and MM acknowledge the NTT PHI Lab for funding. Research at IQC is further supported by the Government of Canada through Innovation, Science and Economic Development Canada (ISED). Research at Perimeter Institute is supported in part by the Government of Canada through ISED and by the Province of Ontario through the Ministry of Colleges and Universities. 

% \newpage
% \quad % without \quad no new page would be insert!!!
% \newpage

% \newpage
% \quad % without \quad no new page would be insert!!!
% \newpage
% \newpage
% \quad % without \quad no new page would be insert!!!
% \newpage

%------------------------------
\appendix
\section*{Appendix}
%\section{Appendix}
\section{Grover's Algorithm}
\label{sec: Grover}
Grover's algorithm is a quantum algorithm to solve the Unstructured Search Problem. From a superposition of all states to be searched over, Grover's algorithm involves successive applications of two operators $U_\psi$ and $U_P$ to ensure that the measurement result at the end of the algorithm gives the search output with high probability. We use this algorithm extensively in our work. The inset of Fig.~\ref{fig:LD_flow} describes the flow of this algorithm for our Local Density computation step. For $m$ points, this involves $O(\sqrt{m})$ successive applications of the operators
\begin{subequations}\label{eq: Grover_operators}
\begin{align}
    U_{\psi} & = 2\ket{\psi}\bra{\psi} - 1
    , \label{eq:Upsi_action}
    \\
    U_P \ket{x} & =
    \begin{cases}
     - \ket{x} & f(x) = 1 \\
     \ket{x} & f(x) = 0
     \end{cases}
     . \label{eq: Up_action}
\end{align}
\end{subequations}
Here, $f(x) = 1$ when the current point satisfies a desired condition (e.g in the context of Local Density Computation, it lies in the critical radius $d_{\rm c}$). If this condition is not satisfied, $f(x) = 0$. 

To implement the operators in Eqs.~\eqref{eq: Grover_operators}, we require A(dd) and M(ultiply) circuits. We use the ones introduced in Ref.~\cite{seidel2021efficient}, which perform the following operations
\begin{subequations}
\label{eq: unitaries}
\begin{align}
    \rm A\ket{X_i} \ket{0}  &= \ket{X_i}\ket{x_{1_i} + x_{2_i}}, \label{eq: U_psi
    }\\
    \rm M \ket{X_i} \ket{0}  &=  \ket{X_i} \ket{x_{1_i} \cdot x_{2_i}} \label{eq: U_P} .
\end{align}
\end{subequations}

\begin{figure} [!htb]
    \centering
    \includegraphics[width=\columnwidth]{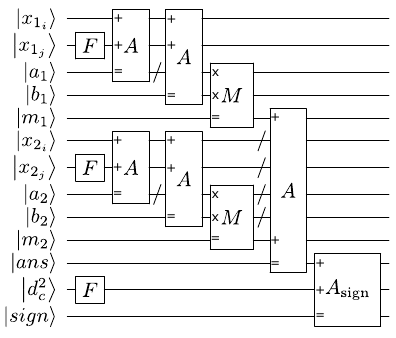}
    \caption{
    Circuit corresponding to the marking operator $U_P$ in Eq.~\eqref{eq: Up_action}. First, it computes the Euclidean distance $d_{1,2}$ between points $1$ and $2$, as defined in Sec.~\ref{subsec: overview}. Then, it compares $d_{1,2}$ to the threshold $d_{\rm c}$ and marks the index based on the sign qubit $\ket{sign}$. $\rm A$ and $\rm B$ are defined in Eq.~\eqref{eq: unitaries} and in Ref.~\cite{seidel2021efficient}. Bars before gates are employed to indicate that the corresponding qubits are unaffected. In the figure, the first subscript on theregister contains the spatial coordinate and the second subscript contains the index. So, for example, $\ket{x_{1_i}}$ contains the $x_1$ coordinate of the $i^{th}$ point. The flip gates $F$ flip the sign bit of the input. The ancillas $a_1$ and $b_1$ contain the output of computation $x_{1_i} - x_{1_j}$, while $a_2$ and $b_2$ contain the output of computation $x_{2_i} - x_{2_j}$, after the first two levels of $A$ gates. The $M$ gates then set $m_1$ and $m_2$ to $(x_{1_i} - x_{1_j})^2$ and $(x_{2_i} - x_{2_j})^2$ respectively. The pre-final $A$ gate sets the $\ket{ans}$ state to $(x_{1_i} - x_{1_j})^2 + (x_{2_i} - x_{2_j})^2$. The final $A_{\rm sign}$ gate computes $(x_{1_i} - x_{1_j})^2 + (x_{2_i} - x_{2_j})^2 - d_c^2$. However, for this final $A_{\rm sign}$ gate, we need only compute the sign qubit of the computation and so only the sign qubit computation subcircuit of $A$ features in $A_{\rm sign}$. 
    }
    \label{fig:blackbox_operator}
\end{figure}

For the local density step, the quantum circuit $U_P$ to implement the search function $f$ is given in Fig.~\ref{fig:blackbox_operator}. The overall idea is to compute the Euclidean distance between every input point and the base point and check if this distance is higher than $d_{\rm c}$. The $\ket{sign}$ qubit stores the output of this computation. The $F$ gates are $X_{sign}$ gates which are $X$ on the sign qubit and act as the identity on every other qubit, such that $F \ket{x_{1_i}} = \ket{-x_{1_i}}$. These are used such that the first and second levels of Add gates $A$ compute $x_{i_1} - x_{i_2}$ into $\ket{a_i}$ and $\ket{b_i}$ respectively. The $M$ multiply gates then set the $\ket{m_i}$ states to $(x_{i_1} - x_{i_2})^2$ taking $a_i$ and $b_i$ as inputs. An $A$ gate next acts on $m_1$ and $m_2$ to set $\ket{ans}$ to $(x_{1_1} - x_{1_2})^2 + (x_{2_1} - x_{2_2})^2$. The $A_{\rm sign}$ gate is a subcircuit of the addition circuit $A$ that finally computes the sign of $(x_{1_1} - x_{1_2})^2 + (x_{2_1} - x_{2_2})^2 - d_c^2$ and stores it in $\ket{sign}$. Thus, $\ket{sign} = \ket{f(x)}$ [with $f(x)$ as in Eq.~\eqref{eq: Up_action}] for the Local Density computation step.

% For the Local Density, the Grover unitary needs to mark points that satisfy a Euclidean distance criterion, concretely, $d_{i,j}<d_{\rm c}$. The description of a quantum circuit for $U_w$ that can perform this marking is provided in Fig.~\ref{fig:blackbox_operator}.

% We can infer then that in Fig.~\ref{fig:blackbox_operator}, $\ket{a_1}$ and $\ket{a_2}$ store $(x_{1_1}-x_{1_2})$ and $(x_{2_1}-x_{2_2})$ respectively. $\ket{m_1}$ and $\ket{m_2}$ store $(x_{1_1} - x_{1_2})^2$ and $(x_{2_1} - x_{2_2})^2$ respectively. $\ket{A}$ stores $||X_1-X_2|| = (x_{1_1}-x_{1_2})^2 + (x_{2_1} - x_{2_2})^2$, the Euclidean distance between points $p_1$ and $p_2$ and $\ket{S}$ stores the sign of $\ket{ans} - d_{\rm c}^2$  and equals the output of the marking function $f$. \\

For the Nearest Higher Procedure, a similar circuit can be used with additional registers $\ket{\rho_i}$ for the Local Density and $\ket{\Tilde{\rho}}$ for the critical density threshold $\Tilde{\rho}$. $d_{\rm c}$ is replaced by $d_{\rm L}$ and the signs of $\ket{d_{i,j}^2} - \ket{ans}$ and $\rho_i - \Tilde{\rho}$ are additionally computed in order to find only the points that lie between $d_{i,j}$ and $d_{\rm L}$ and satisfy the $\Tilde{\rho}$ threshold. These signs can be combined with the Toffoli gate to compute the required function $f(x)$

For Seeds and Outliers, the condition is similar to that in the Nearest Higher (GEBS) procedure in that the distance criterion is only a lower bound as opposed to a window. The GEBS blackbox without the upper distance bound computation can be used.

For the Cluster Assignment step, the operator $U_P$ needs to check if the Nearest Higher index of the points over which Grover Search is carried out lies within $\mathcal{C}$. The $U_P$ operator for this can be generated using $X$ and $C^nZ$ gates. $C^nZ$ gates map $\ket{1}^{\otimes n}$ to $-\ket{1}^{\otimes n}$ and act as the identity on all the other elements of the computational basis.  Let us consider a simple example where the dynamic search space DSS has $8$ points indexed from $000$ to $111$ in binary. So, the Nearest Higher index will always be in this range.  If $\mathcal{C}$ consists of indices $0 (000)$ and $5 (101)$, we would use $U_P$ as defined in Fig.~\ref{fig:find_index}. This flips the phase of $000$ and $101$ in the input superposition as required (following Eq.~\eqref{eq: U_P}). As shown in Fig.~\ref{fig:find_index}, operators for each index can be sequentially applied to check for multiple indices in $\mathcal{C}$.

\begin{figure}[!htb]
    \centering
    \includegraphics[width=0.7\columnwidth]{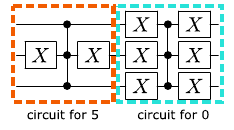}
     \label{fig:find_index}
     \caption{Circuit for $U_P$ to check if the Nearest Higher indices are in $\mathcal{C}$, which here has 2 elements: $5 (101)$ and $0 (000)$. In the above circuit, the first part of the circuit flips the phase of the state if the register is in the $\ket{101}$ state that corresponds to index $5$. The second part flips the phase of the state if the register is in the $\ket{000}$ state that corresponds to index $0$. Note that the $X$ gates performed at the end of the two subcircuits are uncomputation steps that ensure that the input to the subsequent subcircuits are as required and unaffected by our initial $X$ gates.}
     \label{fig:find_index}
\end{figure}

% \section{Grover-Enhanced Binary Search (GEBS)}
% \label{sec: GEBS}
% Grover Search is first used to find points satisfying the condition in Eq.~\eqref{eq: gh}(a). If a point is found, it is first removed from the search space for the next iteration. This can either be done by reinitializing the superposition via qRAM or by modifying the BlackBox Operator to not include points that have already been found. Updates in Eq.~\eqref{eq: gh}(b) are performed. If not, the algorithm enters a branched conditional
% \begin{subequations} \label{eq: gh}
% \begin{align}
%      d_{\rm L} < d_{i,j} < d_{\rm t}, \rho_i < \rho_t, \\
%     d_{t} = (d_{i,j} + d_{\rm L})/2, \rm{\mathcal{N}} = i.
% \end{align}
% \end{subequations}

% If this is the first time the algorithm is entering loop or if $d_{\rm t}$ was just set to $d_{t-1}$, we return $\rm{\mathcal{N}}$. If $d_{\rm t}$ was set to $(d_{i,j} + d_{\rm L})/2$, then the updates in Eq.~\eqref{eq: gh_update1} are performed.

% \begin{equation}
%     d_{\rm t} = d_{t-1}, d_{\rm L} = d_{\rm t}
%     \label{eq: gh_update1}
% \end{equation}

% Fig.~\ref{fig:nh_flow} shows a step-by-step working example of the algorithm to find the Nearest Higher.
\newpage
\printbibliography

@misc{tiwari2020kernel,
      title={Kernel Method based on Non-Linear Coherent State}, 
      author={Prayag Tiwari and Shahram Dehdashti and Abdul Karim Obeid and Massimo Melucci and Peter Bruza},
      year={2020},
      eprint={2007.07887},
      archivePrefix={arXiv},
      primaryClass={quant-ph}
}

@ARTICLE{Zeguendry2023-ss,
  title    = "Quantum Machine Learning: A Review and Case Studies",
  author   = "Zeguendry, Amine and Jarir, Zahi and Quafafou, Mohamed",
  journal  = "Entropy (Basel)",
  volume   =  25,
  number   =  2,
  month    =  feb,
  year     =  2023,
  address  = "Switzerland",
  language = "en"
}

@ARTICLE{Fujita2021-rh,
  title    = "Approximate spectral clustering using both reference vectors and
              topology of the network generated by growing neural gas",
  author   = "Fujita, Kazuhisa",
  journal  = "PeerJ Comput Sci",
  volume   =  7,
  pages    = "e679",
  month    =  aug,
  year     =  2021,
  address  = "United States",
  language = "en"
}

@inproceedings{grover,
author = {Grover, Lov K.},
title = {A Fast Quantum Mechanical Algorithm for Database Search},
year = {1996},
isbn = {0897917855},
publisher = {Association for Computing Machinery},
address = {New York, NY, USA},
url = {https://doi.org/10.1145/237814.237866},
doi = {10.1145/237814.237866},
booktitle = {Proceedings of the Twenty-Eighth Annual ACM Symposium on Theory of Computing},
pages = {212–219},
numpages = {8},
location = {Philadelphia, Pennsylvania, USA},
series = {STOC '96}
}

@techreport{clue_tech,
      author        = "Brondolin, Erica",
      collaboration = "CMS",
      title         = "{CLUE  a clustering algorithm for current and future
                       experiments}",
      institution   = "CERN",
      reportNumber  = "CMS-CR-2022-027",
      address       = "Geneva",
      year          = "2022",
      url           = "https://cds.cern.ch/record/2802590",
}

@misc{cmscollaboration2023development,
      title={Development of the CMS detector for the CERN LHC Run 3}, 
      author={CMS Collaboration},
      year={2023},
      eprint={2309.05466},
      archivePrefix={arXiv},
      primaryClass={physics.ins-det}
}

@ARTICLE{clue,
  
AUTHOR={Rovere, Marco and Chen, Ziheng and Di Pilato, Antonio and Pantaleo, Felice and Seez, Chris},   
	 
TITLE={CLUE: A Fast Parallel Clustering Algorithm for High Granularity Calorimeters in High-Energy Physics},      
	
JOURNAL={Frontiers in Big Data},      
	
VOLUME={3},           
	
YEAR={2020},      
	  
URL={https://www.frontiersin.org/articles/10.3389/fdata.2020.591315},       
	
DOI={10.3389/fdata.2020.591315},      
	
ISSN={2624-909X}
}

@misc{seidel2021efficient,
      title={Efficient Floating Point Arithmetic for Quantum Computers}, 
      author={Raphael Seidel and Nikolay Tcholtchev and Sebastian Bock and Colin Kai-Uwe Becker and Manfred Hauswirth},
      year={2021},
      eprint={2112.10537},
      archivePrefix={arXiv},
      primaryClass={quant-ph}
}

@misc{Brassard_2002,
	doi = {10.1090/conm/305/05215},
  
	url = {https://doi.org/10.1090%2Fconm%2F305%2F05215},
  
	year = 2002,
	publisher = {American Mathematical Society},
  
	pages = {53--74},
  
	author = {Gilles Brassard and Peter H{\o}yer and Michele Mosca and Alain Tapp},
  
	title = {Quantum amplitude amplification and estimation}
}

@article{Giovannetti_2008,
	doi = {10.1103/physrevlett.100.160501},
  
	url = {https://doi.org/10.1103%2Fphysrevlett.100.160501},
  
	year = 2008,
	month = {apr},
  
	publisher = {American Physical Society ({APS})},
  
	volume = {100},
  
	number = {16},
  
	author = {Vittorio Giovannetti and Seth Lloyd and Lorenzo Maccone},
  
	title = {Quantum Random Access Memory},
  
	journal = {Physical Review Letters}
}

@misc{nicotra2023quantum,
      title={A quantum algorithm for track reconstruction in the LHCb vertex detector}, 
      author={Davide Nicotra and Miriam Lucio Martinez and Jacco Andreas de Vries and Marcel Merk and Kurt Driessens and Ronald Leonard Westra and Domenica Dibenedetto and Daniel Hugo Cámpora Pérez},
      year={2023},
      eprint={2308.00619},
      archivePrefix={arXiv},
      primaryClass={quant-ph}
}

@Article{Labuhn2016,
author={Labuhn, Henning
and Barredo, Daniel
and Ravets, Sylvain
and de L{\'e}s{\'e}leuc, Sylvain
and Macr{\`i}, Tommaso
and Lahaye, Thierry
and Browaeys, Antoine},
title={Tunable two-dimensional arrays of single Rydberg atoms for realizing quantum Ising models},
journal={Nature},
year={2016},
month={Jun},
day={01},
volume={534},
number={7609},
pages={667-670},
issn={1476-4687},
doi={10.1038/nature18274},
url={https://doi.org/10.1038/nature18274}
}

@article{Lanyon2011,
author = {B. P. Lanyon  and C. Hempel  and D. Nigg  and M. Müller  and R. Gerritsma  and F. Zähringer  and P. Schindler  and J. T. Barreiro  and M. Rambach  and G. Kirchmair  and M. Hennrich  and P. Zoller  and R. Blatt  and C. F. Roos },
title = {Universal Digital Quantum Simulation with Trapped Ions},
journal = {Science},
volume = {334},
number = {6052},
pages = {57-61},
year = {2011},
doi = {10.1126/science.1208001},
URL = {https://www.science.org/doi/abs/10.1126/science.1208001},
eprint = {https://www.science.org/doi/pdf/10.1126/science.1208001}}

@Article{Debnath2016,
author={Debnath, S.
and Linke, N. M.
and Figgatt, C.
and Landsman, K. A.
and Wright, K.
and Monroe, C.},
title={Demonstration of a small programmable quantum computer with atomic qubits},
journal={Nature},
year={2016},
month={Aug},
day={01},
volume={536},
number={7614},
pages={63-66},
issn={1476-4687},
doi={10.1038/nature18648},
url={https://doi.org/10.1038/nature18648}
}

@Article{Córcoles2015,
author={C{\'o}rcoles, A. D.
and Magesan, Easwar
and Srinivasan, Srikanth J.
and Cross, Andrew W.
and Steffen, M.
and Gambetta, Jay M.
and Chow, Jerry M.},
title={Demonstration of a quantum error detection code using a square lattice of four superconducting qubits},
journal={Nature Communications},
year={2015},
month={Apr},
day={29},
volume={6},
number={1},
pages={6979},
issn={2041-1723},
doi={10.1038/ncomms7979},
url={https://doi.org/10.1038/ncomms7979}
}

@Article{Arute2019,
author={Arute, Frank
and Arya, Kunal
and Babbush, Ryan
and Bacon, Dave
and Bardin, Joseph C.
and Barends, Rami
and Biswas, Rupak
and Boixo, Sergio
and Brandao, Fernando G. S. L.
and Buell, David A.
and Burkett, Brian
and Chen, Yu
and Chen, Zijun
and Chiaro, Ben
and Collins, Roberto
and Courtney, William
and Dunsworth, Andrew
and Farhi, Edward
and Foxen, Brooks
and Fowler, Austin
and Gidney, Craig
and Giustina, Marissa
and Graff, Rob
and Guerin, Keith
and Habegger, Steve
and Harrigan, Matthew P.
and Hartmann, Michael J.
and Ho, Alan
and Hoffmann, Markus
and Huang, Trent
and Humble, Travis S.
and Isakov, Sergei V.
and Jeffrey, Evan
and Jiang, Zhang
and Kafri, Dvir
and Kechedzhi, Kostyantyn
and Kelly, Julian
and Klimov, Paul V.
and Knysh, Sergey
and Korotkov, Alexander
and Kostritsa, Fedor
and Landhuis, David
and Lindmark, Mike
and Lucero, Erik
and Lyakh, Dmitry
and Mandr{\`a}, Salvatore
and McClean, Jarrod R.
and McEwen, Matthew
and Megrant, Anthony
and Mi, Xiao
and Michielsen, Kristel
and Mohseni, Masoud
and Mutus, Josh
and Naaman, Ofer
and Neeley, Matthew
and Neill, Charles
and Niu, Murphy Yuezhen
and Ostby, Eric
and Petukhov, Andre
and Platt, John C.
and Quintana, Chris
and Rieffel, Eleanor G.
and Roushan, Pedram
and Rubin, Nicholas C.
and Sank, Daniel
and Satzinger, Kevin J.
and Smelyanskiy, Vadim
and Sung, Kevin J.
and Trevithick, Matthew D.
and Vainsencher, Amit
and Villalonga, Benjamin
and White, Theodore
and Yao, Z. Jamie
and Yeh, Ping
and Zalcman, Adam
and Neven, Hartmut
and Martinis, John M.},
title={Quantum supremacy using a programmable superconducting processor},
journal={Nature},
year={2019},
month={Oct},
day={01},
volume={574},
number={7779},
pages={505-510},
issn={1476-4687},
doi={10.1038/s41586-019-1666-5},
url={https://doi.org/10.1038/s41586-019-1666-5}
}

@Article{Bernien2017,
author={Bernien, Hannes
and Schwartz, Sylvain
and Keesling, Alexander
and Levine, Harry
and Omran, Ahmed
and Pichler, Hannes
and Choi, Soonwon
and Zibrov, Alexander S.
and Endres, Manuel
and Greiner, Markus
and Vuleti{\'{c}}, Vladan
and Lukin, Mikhail D.},
title={Probing many-body dynamics on a 51-atom quantum simulator},
journal={Nature},
year={2017},
month={Nov},
day={01},
volume={551},
number={7682},
pages={579-584},
issn={1476-4687},
doi={10.1038/nature24622},
url={https://doi.org/10.1038/nature24622}
}

@article{tetraquark,
  title = {First Observation of a Doubly Charged Tetraquark and Its Neutral Partner},
  author = {Aaij, R. and Abdelmotteleb, A. S. W. and Abellan Beteta, C. and Abudin\'en, F. and Ackernley, T. and Adeva, B. and Adinolfi, M. and Adlarson, P. and Afsharnia, H. and Agapopoulou, C. and Aidala, C. A. and Aiola, S. and Ajaltouni, Z. and Akar, S. and Akiba, K. and Albrecht, J. and Alessio, F. and Alexander, M. and Alfonso Albero, A. and Aliouche, Z. and Alvarez Cartelle, P. and Amalric, R. and Amato, S. and Amey, J. L. and Amhis, Y. and An, L. and Anderlini, L. and Andersson, M. and Andreianov, A. and Andreotti, M. and Andreou, D. and Ao, D. and Archilli, F. and Artamonov, A. and Artuso, M. and Aslanides, E. and Atzeni, M. and Audurier, B. and Bachmann, S. and Bachmayer, M. and Back, J. J. and Bailly-reyre, A. and Baladron Rodriguez, P. and Balagura, V. and Baldini, W. and Baptista de Souza Leite, J. and Barbetti, M. and Barlow, R. J. and Barsuk, S. and Barter, W. and Bartolini, M. and Baryshnikov, F. and Basels, J. M. and Bassi, G. and Batsukh, B. and Battig, A. and Bay, A. and Beck, A. and Becker, M. and Bedeschi, F. and Bediaga, I. B. and Beiter, A. and Belavin, V. and Belin, S. and Bellee, V. and Belous, K. and Belov, I. and Belyaev, I. and Benane, G. and Bencivenni, G. and Ben-Haim, E. and Berezhnoy, A. and Bernet, R. and Bernet Andres, S. and Berninghoff, D. and Bernstein, H. C. and Bertella, C. and Bertolin, A. and Betancourt, C. and Betti, F. and Bezshyiko, Ia. and Bhasin, S. and Bhom, J. and Bian, L. and Bieker, M. S. and Biesuz, N. V. and Bifani, S. and Billoir, P. and Biolchini, A. and Birch, M. and Bishop, F. C. R. and Bitadze, A. and Bizzeti, A. and Blago, M. P. and Blake, T. and Blanc, F. and Blank, J. E. and Blusk, S. and Bobulska, D. and Boelhauve, J. A. and Boente Garcia, O. and Boettcher, T. and Boldyrev, A. and Bolognani, C. S. and Bolzonella, R. and Bondar, N. and Borgato, F. and Borghi, S. and Borsato, M. and Borsuk, J. T. and Bouchiba, S. A. and Bowcock, T. J. V. and Boyer, A. and Bozzi, C. and Bradley, M. J. and Braun, S. and Brea Rodriguez, A. and Brodzicka, J. and Brossa Gonzalo, A. and Brown, J. and Brundu, D. and Buonaura, A. and Buonincontri, L. and Burke, A. T. and Burr, C. and Bursche, A. and Butkevich, A. and Butter, J. S. and Buytaert, J. and Byczynski, W. and Cadeddu, S. and Cai, H. and Calabrese, R. and Calefice, L. and Cali, S. and Calladine, R. and Calvi, M. and Calvo Gomez, M. and Campana, P. and Campora Perez, D. H. and Campoverde Quezada, A. F. and Capelli, S. and Capriotti, L. and Carbone, A. and Carboni, G. and Cardinale, R. and Cardini, A. and Carniti, P. and Carus, L. and Casais Vidal, A. and Caspary, R. and Casse, G. and Cattaneo, M. and Cavallero, G. and Cavallini, V. and Celani, S. and Cerasoli, J. and Cervenkov, D. and Chadwick, A. J. and Chapman, M. G. and Charles, M. and Charpentier, Ph. and Chavez Barajas, C. A. and Chefdeville, M. and Chen, C. and Chen, S. and Chernov, A. and Chernyshenko, S. and Chobanova, V. and Cholak, S. and Chrzaszcz, M. and Chubykin, A. and Chulikov, V. and Ciambrone, P. and Cicala, M. F. and Cid Vidal, X. and Ciezarek, G. and Ciullo, G. and Clarke, P. E. L. and Clemencic, M. and Cliff, H. V. and Closier, J. and Cobbledick, J. L. and Coco, V. and Coelho, J. A. B. and Cogan, J. and Cogneras, E. and Cojocariu, L. and Collins, P. and Colombo, T. and Congedo, L. and Contu, A. and Cooke, N. and Corredoira, I. and Corti, G. and Couturier, B. and Craik, D. C. and Cruz Torres, M. and Currie, R. and Da Silva, C. L. and Dadabaev, S. and Dai, L. and Dai, X. and Dall'Occo, E. and Dalseno, J. and D'Ambrosio, C. and Daniel, J. and Danilina, A. and d'Argent, P. and Davies, J. E. and Davis, A. and De Aguiar Francisco, O. and de Boer, J. and De Bruyn, K. and De Capua, S. and De Cian, M. and De Freitas Carneiro Da Graca, U. and De Lucia, E. and De Miranda, J. M. and De Paula, L. and De Serio, M. and De Simone, D. and De Simone, P. and De Vellis, F. and de Vries, J. A. and Dean, C. T. and Debernardis, F. and Decamp, D. and Dedu, V. and Del Buono, L. and Delaney, B. and Dembinski, H.-P. and Denysenko, V. and Deschamps, O. and Dettori, F. and Dey, B. and Di Cicco, A. and Di Nezza, P. and Diachkov, I. and Didenko, S. and Dieste Maronas, L. and Ding, S. and Dobishuk, V. and Dolmatov, A. and Dong, C. and Donohoe, A. M. and Dordei, F. and dos Reis, A. C. and Douglas, L. and Downes, A. G. and Duda, P. and Dudek, M. W. and Dufour, L. and Duk, V. and Durante, P. and Duras, M. M. and Durham, J. M. and Dutta, D. and Dziurda, A. and Dzyuba, A. and Easo, S. and Egede, U. and Egorychev, V. and Eidelman, S. and Eirea Orro, C. and Eisenhardt, S. and Ejopu, E. and Ek-In, S. and Eklund, L. and Ely, S. and Ene, A. and Epple, E. and Escher, S. and Eschle, J. and Esen, S. and Evans, T. and Fabiano, F. and Falcao, L. N. and Fan, Y. and Fang, B. and Fantini, L. and Faria, M. and Farry, S. and Fazzini, D. and Felkowski, L. F and Feo, M. and Fernandez Gomez, M. and Fernez, A. D. and Ferrari, F. and Ferreira Lopes, L. and Ferreira Rodrigues, F. and Ferreres Sole, S. and Ferrillo, M. and Ferro-Luzzi, M. and Filippov, S. and Fini, R. A. and Fiorini, M. and Firlej, M. and Fischer, K. M. and Fitzgerald, D. S. and Fitzpatrick, C. and Fiutowski, T. and Fleuret, F. and Fontana, M. and Fontanelli, F. and Forty, R. and Foulds-Holt, D. and Franco Lima, V. and Franco Sevilla, M. and Frank, M. and Franzoso, E. and Frau, G. and Frei, C. and Friday, D. A. and Fu, J. and Fuehring, Q. and Fulghesu, T. and Gabriel, E. and Galati, G. and Galati, M. D. and Gallas Torreira, A. and Galli, D. and Gambetta, S. and Gan, Y. and Gandelman, M. and Gandini, P. and Gao, Y. and Gao, Y. and Garau, M. and Garcia Martin, L. M. and Garcia Moreno, P. and Garc\'{\i}a Pardi\~nas, J. and Garcia Plana, B. and Garcia Rosales, F. A. and Garrido, L. and Gaspar, C. and Geertsema, R. E. and Gerick, D. and Gerken, L. L. and Gersabeck, E. and Gersabeck, M. and Gershon, T. and Giambastiani, L. and Gibson, V. and Giemza, H. K. and Gilman, A. L. and Giovannetti, M. and Giovent\`u, A. and Gironella Gironell, P. and Giugliano, C. and Giza, M. A. and Gizdov, K. and Gkougkousis, E. L. and Gligorov, V. V. and G\"obel, C. and Golobardes, E. and Golubkov, D. and Golutvin, A. and Gomes, A. and Gomez Fernandez, S. and Goncalves Abrantes, F. and Goncerz, M. and Gong, G. and Gorelov, I. V. and Gotti, C. and Grabowski, J. P. and Grammatico, T. and Granado Cardoso, L. A. and Graug\'es, E. and Graverini, E. and Graziani, G. and Grecu, A. T. and Greeven, L. M. and Grieser, N. A. and Grillo, L. and Gromov, S. and Gruberg Cazon, B. R. and Gu, C. and Guarise, M. and Guittiere, M. and G\"unther, P. A. and Gushchin, E. and Guth, A. and Guz, Y. and Gys, T. and Hadavizadeh, T. and Haefeli, G. and Haen, C. and Haimberger, J. and Haines, S. C. and Halewood-leagas, T. and Halvorsen, M. M. and Hamilton, P. M. and Hammerich, J. and Han, Q. and Han, X. and Hansen, E. B. and Hansmann-Menzemer, S. and Hao, L. and Harnew, N. and Harrison, T. and Hasse, C. and Hatch, M. and He, J. and Heijhoff, K. and Henderson, C. and Henderson, R. D. L. and Hennequin, A. M. and Hennessy, K. and Henry, L. and Herd, J. and Heuel, J. and Hicheur, A. and Hill, D. and Hilton, M. and Hollitt, S. E. and Horswill, J. and Hou, R. and Hou, Y. and Hu, J. and Hu, J. and Hu, W. and Hu, X. and Huang, W. and Huang, X. and Hulsbergen, W. and Hunter, R. J. and Hushchyn, M. and Hutchcroft, D. and Ibis, P. and Idzik, M. and Ilin, D. and Ilten, P. and Inglessi, A. and Iniukhin, A. and Ishteev, A. and Ivshin, K. and Jacobsson, R. and Jage, H. and Jaimes Elles, S. J. and Jakobsen, S. and Jans, E. and Jashal, B. K. and Jawahery, A. and Jevtic, V. and Jiang, E. and Jiang, X. and Jiang, Y. and John, M. and Johnson, D. and Jones, C. R. and Jones, T. P. and Jost, B. and Jurik, N. and Juszczak, I. and Kandybei, S. and Kang, Y. and Karacson, M. and Karpenkov, D. and Karpov, M. and Kautz, J. W. and Keizer, F. and Keller, D. M. and Kenzie, M. and Ketel, T. and Khanji, B. and Kharisova, A. and Kholodenko, S. and Khreich, G. and Kirn, T. and Kirsebom, V. S. and Kitouni, O. and Klaver, S. and Kleijne, N. and Klimaszewski, K. and Kmiec, M. R. and Koliiev, S. and Kondybayeva, A. and Konoplyannikov, A. and Kopciewicz, P. and Kopecna, R. and Koppenburg, P. and Korolev, M. and Kostiuk, I. and Kot, O. and Kotriakhova, S. and Kozachuk, A. and Kravchenko, P. and Kravchuk, L. and Krawczyk, R. D. and Kreps, M. and Kretzschmar, S. and Krokovny, P. and Krupa, W. and Krzemien, W. and Kubat, J. and Kubis, S. and Kucewicz, W. and Kucharczyk, M. and Kudryavtsev, V. and Kupsc, A. and Lacarrere, D. and Lafferty, G. and Lai, A. and Lampis, A. and Lancierini, D. and Landesa Gomez, C. and Lane, J. J. and Lane, R. and Lanfranchi, G. and Langenbruch, C. and Langer, J. and Lantwin, O. and Latham, T. and Lazzari, F. and Lazzaroni, M. and Le Gac, R. and Lee, S. H. and Lef\`evre, R. and Leflat, A. and Legotin, S. and Lenisa, P. and Leroy, O. and Lesiak, T. and Leverington, B. and Li, A. and Li, H. and Li, K. and Li, P. and Li, P.-R. and Li, S. and Li, T. and Li, T. and Li, Y. and Li, Z. and Liang, X. and Lin, C. and Lin, T. and Lindner, R. and Lisovskyi, V. and Litvinov, R. and Liu, G. and Liu, H. and Liu, Q. and Liu, S. and Liu, Y. and Lobo Salvia, A. and Loi, A. and Lollini, R. and Lomba Castro, J. and Longstaff, I. and Lopes, J. H. and Lopez Huertas, A. and L\'opez Soli\~no, S. and Lovell, G. H. and Lu, Y. and Lucarelli, C. and Lucchesi, D. and Luchuk, S. and Lucio Martinez, M. and Lukashenko, V. and Luo, Y. and Lupato, A. and Luppi, E. and Lusiani, A. and Lynch, K. and Lyu, X.-R. and Ma, L. and Ma, R. and Maccolini, S. and Machefert, F. and Maciuc, F. and Mackay, I. and Macko, V. and Mackowiak, P. and Madhan Mohan, L. R. and Maevskiy, A. and Maisuzenko, D. and Majewski, M. W. and Malczewski, J. J. and Malde, S. and Malecki, B. and Malinin, A. and Maltsev, T. and Manca, G. and Mancinelli, G. and Mancuso, C. and Manuzzi, D. and Manzari, C. A. and Marangotto, D. and Marchand, J. F. and Marconi, U. and Mariani, S. and Marin Benito, C. and Marks, J. and Marshall, A. M. and Marshall, P. J. and Martelli, G. and Martellotti, G. and Martinazzoli, L. and Martinelli, M. and Martinez Santos, D. and Martinez Vidal, F. and Massafferri, A. and Materok, M. and Matev, R. and Mathad, A. and Matiunin, V. and Matteuzzi, C. and Mattioli, K. R. and Mauri, A. and Maurice, E. and Mauricio, J. and Mazurek, M. and McCann, M. and Mcconnell, L. and McGrath, T. H. and McHugh, N. T. and McNab, A. and McNulty, R. and Mead, J. V. and Meadows, B. and Meier, G. and Melnychuk, D. and Meloni, S. and Merk, M. and Merli, A. and Meyer Garcia, L. and Miao, D. and Mikhasenko, M. and Milanes, D. A. and Millard, E. and Milovanovic, M. and Minard, M.-N. and Minotti, A. and Miralles, T. and Mitchell, S. E. and Mitreska, B. and Mitzel, D. S. and M\"odden, A. and Mohammed, R. A. and Moise, R. D. and Mokhnenko, S. and Momb\"acher, T. and Monk, M. and Monroy, I. A. and Monteil, S. and Morandin, M. and Morello, G. and Morello, M. J. and Moron, J. and Morris, A. B. and Morris, A. G. and Mountain, R. and Mu, H. and Muhammad, E. and Muheim, F. and Mulder, M. and M\"uller, K. and Murphy, C. H. and Murray, D. and Murta, R. and Muzzetto, P. and Naik, P. and Nakada, T. and Nandakumar, R. and Nanut, T. and Nasteva, I. and Needham, M. and Neri, N. and Neubert, S. and Neufeld, N. and Neustroev, P. and Newcombe, R. and Nicolini, J. and Niel, E. M. and Nieswand, S. and Nikitin, N. and Nolte, N. S. and Normand, C. and Novoa Fernandez, J. and Nunez, C. and Oblakowska-Mucha, A. and Obraztsov, V. and Oeser, T. and O'Hanlon, D. P. and Okamura, S. and Oldeman, R. and Oliva, F. and Onderwater, C. J. G. and O'Neil, R. H. and Otalora Goicochea, J. M. and Ovsiannikova, T. and Owen, P. and Oyanguren, A. and Ozcelik, O. and Padeken, K. O. and Pagare, B. and Pais, P. R. and Pajero, T. and Palano, A. and Palutan, M. and Pan, Y. and Panshin, G. and Paolucci, L. and Papanestis, A. and Pappagallo, M. and Pappalardo, L. L. and Pappenheimer, C. and Parker, W. and Parkes, C. and Passalacqua, B. and Passaleva, G. and Pastore, A. and Patel, M. and Patrignani, C. and Pawley, C. J. and Pearce, A. and Pellegrino, A. and Pepe Altarelli, M. and Perazzini, S. and Pereima, D. and Pereiro Castro, A. and Perret, P. and Petric, M. and Petridis, K. and Petrolini, A. and Petrov, A. and Petrucci, S. and Petruzzo, M. and Pham, H. and Philippov, A. and Piandani, R. and Pica, L. and Piccini, M. and Pietrzyk, B. and Pietrzyk, G. and Pili, M. and Pinci, D. and Pisani, F. and Pizzichemi, M. and Placinta, V. and Plews, J. and Plo Casasus, M. and Polci, F. and Poli Lener, M. and Poliakova, M. and Poluektov, A. and Polukhina, N. and Polyakov, I. and Polycarpo, E. and Ponce, S. and Popov, D. and Popov, S. and Poslavskii, S. and Prasanth, K. and Promberger, L. and Prouve, C. and Pugatch, V. and Puill, V. and Punzi, G. and Qi, H. R. and Qian, W. and Qin, N. and Qu, S. and Quagliani, R. and Raab, N. V. and Rabadan Trejo, R. I. and Rachwal, B. and Rademacker, J. H. and Rajagopalan, R. and Rama, M. and Ramos Pernas, M. and Rangel, M. S. and Ratnikov, F. and Raven, G. and Rebollo De Miguel, M. and Redi, F. and Reich, J. and Reiss, F. and Remon Alepuz, C. and Ren, Z. and Resmi, P. K. and Ribatti, R. and Ricci, A. M. and Ricciardi, S. and Richardson, K. and Richardson-Slipper, M. and Rinnert, K. and Robbe, P. and Robertson, G. and Rodrigues, A. B. and Rodrigues, E. and Rodriguez Fernandez, E. and Rodriguez Lopez, J. A. and Rodriguez Rodriguez, E. and Rolf, D. L. and Rollings, A. and Roloff, P. and Romanovskiy, V. and Romero Lamas, M. and Romero Vidal, A. and Roth, J. D. and Rotondo, M. and Rudolph, M. S. and Ruf, T. and Ruiz Fernandez, R. A. and Ruiz Vidal, J. and Ryzhikov, A. and Ryzka, J. and Saborido Silva, J. J. and Sagidova, N. and Sahoo, N. and Saitta, B. and Salomoni, M. and Sanchez Gras, C. and Sanderswood, I. and Santacesaria, R. and Santamarina Rios, C. and Santimaria, M. and Santovetti, E. and Saranin, D. and Sarpis, G. and Sarpis, M. and Sarti, A. and Satriano, C. and Satta, A. and Saur, M. and Savrina, D. and Sazak, H. and Scantlebury Smead, L. G. and Scarabotto, A. and Schael, S. and Scherl, S. and Schiller, M. and Schindler, H. and Schmelling, M. and Schmidt, B. and Schmitt, S. and Schneider, O. and Schopper, A. and Schubiger, M. and Schulte, S. and Schune, M. H. and Schwemmer, R. and Sciascia, B. and Sciuccati, A. and Sellam, S. and Semennikov, A. and Senghi Soares, M. and Sergi, A. and Serra, N. and Sestini, L. and Seuthe, A. and Shang, Y. and Shangase, D. M. and Shapkin, M. and Shchemerov, I. and Shchutska, L. and Shears, T. and Shekhtman, L. and Shen, Z. and Sheng, S. and Shevchenko, V. and Shi, B. and Shields, E. B. and Shimizu, Y. and Shmanin, E. and Shorkin, R. and Shupperd, J. D. and Siddi, B. G. and Silva Coutinho, R. and Simi, G. and Simone, S. and Singla, M. and Skidmore, N. and Skuza, R. and Skwarnicki, T. and Slater, M. W. and Smallwood, J. C. and Smeaton, J. G. and Smith, E. and Smith, K. and Smith, M. and Snoch, A. and Soares Lavra, L. and Sokoloff, M. D. and Soler, F. J. P. and Solomin, A. and Solovev, A. and Solovyev, I. and Song, R. and Souza De Almeida, F. L. and Souza De Paula, B. and Spaan, B. and Spadaro Norella, E. and Spedicato, E. and Spiridenkov, E. and Spradlin, P. and Sriskaran, V. and Stagni, F. and Stahl, M. and Stahl, S. and Stanislaus, S. and Stein, E. N. and Steinkamp, O. and Stenyakin, O. and Stevens, H. and Stone, S. and Strekalina, D. and Suljik, F. and Sun, J. and Sun, L. and Sun, Y. and Svihra, P. and Swallow, P. N. and Swientek, K. and Szabelski, A. and Szumlak, T. and Szymanski, M. and Tan, Y. and Taneja, S. and Tanner, A. R. and Tat, M. D. and Terentev, A. and Teubert, F. and Thomas, E. and Thompson, D. J. D. and Thomson, K. A. and Tilquin, H. and Tisserand, V. and T'Jampens, S. and Tobin, M. and Tomassetti, L. and Tonani, G. and Tong, X. and Torres Machado, D. and Tou, D. Y. and Trilov, S. M. and Trippl, C. and Tuci, G. and Tully, A. and Tuning, N. and Ukleja, A. and Unverzagt, D. J. and Usachov, A. and Ustyuzhanin, A. and Uwer, U. and Vagner, A. and Vagnoni, V. and Valassi, A. and Valenti, G. and Valls Canudas, N. and van Beuzekom, M. and Van Dijk, M. and Van Hecke, H. and van Herwijnen, E. and Van Hulse, C. B. and van Veghel, M. and Vazquez Gomez, R. and Vazquez Regueiro, P. and V\'azquez Sierra, C. and Vecchi, S. and Velthuis, J. J. and Veltri, M. and Venkateswaran, A. and Veronesi, M. and Vesterinen, M. and Vieira, D. and Vieites Diaz, M. and Vilasis-Cardona, X. and Vilella Figueras, E. and Villa, A. and Vincent, P. and Volle, F. C. and vom Bruch, D. and Vorobyev, A. and Vorobyev, V. and Voropaev, N. and Vos, K. and Vrahas, C. and Waldi, R. and Walsh, J. and Wan, G. and Wang, C. and Wang, G. and Wang, J. and Wang, J. and Wang, J. and Wang, J. and Wang, M. and Wang, R. and Wang, X. and Wang, Y. and Wang, Z. and Wang, Z. and Wang, Z. and Ward, J. A. and Watson, N. K. and Websdale, D. and Wei, Y. and Weisser, C. and Westhenry, B. D. C. and White, D. J. and Whitehead, M. and Wiederhold, A. R. and Wiedner, D. and Wilkinson, G. and Wilkinson, M. K. and Williams, I. and Williams, M. and Williams, M. R. J. and Williams, R. and Wilson, F. F. and Wislicki, W. and Witek, M. and Witola, L. and Wong, C. P. and Wormser, G. and Wotton, S. A. and Wu, H. and Wu, J. and Wyllie, K. and Xiang, Z. and Xiao, D. and Xie, Y. and Xu, A. and Xu, J. and Xu, L. and Xu, L. and Xu, M. and Xu, Q. and Xu, Z. and Xu, Z. and Yang, D. and Yang, S. and Yang, X. and Yang, Y. and Yang, Z. and Yang, Z. and Yeomans, L. E. and Yeroshenko, V. and Yeung, H. and Yin, H. and Yu, J. and Yuan, X. and Zaffaroni, E. and Zavertyaev, M. and Zdybal, M. and Zenaiev, O. and Zeng, M. and Zhang, C. and Zhang, D. and Zhang, L. and Zhang, S. and Zhang, S. and Zhang, Y. and Zhang, Y. and Zharkova, A. and Zhelezov, A. and Zheng, Y. and Zhou, T. and Zhou, X. and Zhou, Y. and Zhovkovska, V. and Zhu, X. and Zhu, X. and Zhu, Z. and Zhukov, V. and Zou, Q. and Zucchelli, S. and Zuliani, D. and Zunica, G.},
  collaboration = {LHCb Collaboration},
  journal = {Phys. Rev. Lett.},
  volume = {131},
  issue = {4},
  pages = {041902},
  numpages = {12},
  year = {2023},
  month = {Jul},
  publisher = {American Physical Society},
  doi = {10.1103/PhysRevLett.131.041902},
  url = {https://link.aps.org/doi/10.1103/PhysRevLett.131.041902}
}

@INPROCEEDINGS{astrophysics,
       author = {{Galluccio}, L. and {Michel}, O. and {Bendjoya}, P. and {Slezak}, E.},
        title = "{Unsupervised Clustering on Astrophysics Data: Asteroids Reflectance Spectra Surveys and Hyperspectral Images}",
    booktitle = {Classification and Discovery in Large Astronomical Surveys},
         year = 2008,
       editor = {{Bailer-Jones}, Coryn A.~L.},
       series = {American Institute of Physics Conference Series},
       volume = {1082},
        month = dec,
        pages = {165-171},
          doi = {10.1063/1.3059034},
       adsurl = {https://ui.adsabs.harvard.edu/abs/2008AIPC.1082..165G}
}

@article{med_image,
    author = {Qaqish, Bahjat F and O’Brien, Jonathon J and Hibbard, Jonathan C and Clowers, Katie J},
    title = "{Accelerating high-dimensional clustering with lossless data reduction}",
    journal = {Bioinformatics},
    volume = {33},
    number = {18},
    pages = {2867-2872},
    year = {2017},
    month = {05},
    issn = {1367-4803},
    doi = {10.1093/bioinformatics/btx328},
    url = {https://doi.org/10.1093/bioinformatics/btx328},
    eprint = {https://academic.oup.com/bioinformatics/article-pdf/33/18/2867/49040691/bioinformatics\_33\_18\_2867.pdf},
}

@article{T_ys_z_2020,
   title={Particle Track Reconstruction with Quantum Algorithms},
   volume={245},
   ISSN={2100-014X},
   url={http://dx.doi.org/10.1051/epjconf/202024509013},
   DOI={10.1051/epjconf/202024509013},
   journal={EPJ Web of Conferences},
   publisher={EDP Sciences},
   author={Tüysüz, Cenk and Carminati, Federico and Demirköz, Bilge and Dobos, Daniel and Fracas, Fabio and Novotny, Kristiane and Potamianos, Karolos and Vallecorsa, Sofia and Vlimant, Jean-Roch},
   editor={Doglioni, C. and Kim, D. and Stewart, G.A. and Silvestris, L. and Jackson, P. and Kamleh, W.},
   year={2020},
   pages={09013} }

@inproceedings{quantum_clustering,
author = {A\"{\i}meur, Esma and Brassard, Gilles and Gambs, S\'{e}bastien},
title = {Quantum Clustering Algorithms},
year = {2007},
isbn = {9781595937933},
publisher = {Association for Computing Machinery},
address = {New York, NY, USA},
url = {https://doi.org/10.1145/1273496.1273497},
doi = {10.1145/1273496.1273497},
booktitle = {Proceedings of the 24th International Conference on Machine Learning},
pages = {1–8},
numpages = {8},
location = {Corvalis, Oregon, USA},
series = {ICML '07}
}

@misc{pedregosa2018scikitlearn,
      title={Scikit-learn: Machine Learning in Python}, 
      author={Fabian Pedregosa and Gaël Varoquaux and Alexandre Gramfort and Vincent Michel and Bertrand Thirion and Olivier Grisel and Mathieu Blondel and Andreas Müller and Joel Nothman and Gilles Louppe and Peter Prettenhofer and Ron Weiss and Vincent Dubourg and Jake Vanderplas and Alexandre Passos and David Cournapeau and Matthieu Brucher and Matthieu Perrot and Édouard Duchesnay},
      year={2018},
      eprint={1201.0490},
      archivePrefix={arXiv},
      primaryClass={cs.LG}
}

@article{markets,
title = {Market segmentation using high-dimensional sparse consumers data},
journal = {Expert Systems with Applications},
volume = {145},
pages = {113136},
year = {2020},
issn = {0957-4174},
doi = {https://doi.org/10.1016/j.eswa.2019.113136},
url = {https://www.sciencedirect.com/science/article/pii/S095741741930853X},
author = {Jian Zhou and Linli Zhai and Athanasios A. Pantelous}
}

@article{bioinf,
    author = {Karim, Md Rezaul and Beyan, Oya and Zappa, Achille and Costa, Ivan G and Rebholz-Schuhmann, Dietrich and Cochez, Michael and Decker, Stefan},
    title = "{Deep learning-based clustering approaches for bioinformatics}",
    journal = {Briefings in Bioinformatics},
    volume = {22},
    number = {1},
    pages = {393-415},
    year = {2020},
    month = {02},
    issn = {1477-4054},
    doi = {10.1093/bib/bbz170},
    url = {https://doi.org/10.1093/bib/bbz170},
    eprint = {https://academic.oup.com/bib/article-pdf/22/1/393/35934885/bbz170.pdf},
}

@misc{pires2021digital,
      title={A Digital Quantum Algorithm for Jet Clustering in High-Energy Physics}, 
      author={Diogo Pires and Pedrame Bargassa and João Seixas and Yasser Omar},
      year={2021},
      eprint={2101.05618},
      archivePrefix={arXiv},
      primaryClass={physics.data-an}
}

@article{WU201926,
title = {Hyperparameter Optimization for Machine Learning Models Based on Bayesian Optimizationb},
journal = {Journal of Electronic Science and Technology},
volume = {17},
number = {1},
pages = {26-40},
year = {2019},
issn = {1674-862X},
doi = {https://doi.org/10.11989/JEST.1674-862X.80904120},
url = {https://www.sciencedirect.com/science/article/pii/S1674862X19300047},
author = {Jia Wu and Xiu-Yun Chen and Hao Zhang and Li-Dong Xiong and Hang Lei and Si-Hao Deng}
}

@article{lukin,
  title = {Emerging Two-Dimensional Gauge Theories in Rydberg Configurable Arrays},
  author = {Celi, Alessio and Vermersch, Beno\^{\i}t and Viyuela, Oscar and Pichler, Hannes and Lukin, Mikhail D. and Zoller, Peter},
  journal = {Phys. Rev. X},
  volume = {10},
  issue = {2},
  pages = {021057},
  numpages = {18},
  year = {2020},
  month = {Jun},
  publisher = {American Physical Society},
  doi = {10.1103/PhysRevX.10.021057},
  url = {https://link.aps.org/doi/10.1103/PhysRevX.10.021057}
}

@article{simplifyenrichment,
    author = {Gu, Zuguang and Hübschmann, Daniel},
    title = "{SimplifyEnrichment: A Bioconductor Package for Clustering and Visualizing Functional Enrichment Results}",
    journal = {Genomics, Proteomics and Bioinformatics},
    volume = {21},
    number = {1},
    pages = {190-202},
    year = {2022},
    month = {06},
    issn = {1672-0229},
    doi = {10.1016/j.gpb.2022.04.008},
    url = {https://doi.org/10.1016/j.gpb.2022.04.008},
    eprint = {https://academic.oup.com/gpb/article-pdf/21/1/190/51559608/gpb\_21\_1\_190.pdf},
}

@INPROCEEDINGS{clusteringapps,
  author={Oyelade, Jelili and Isewon, Itunuoluwa and Oladipupo, Olufunke and Emebo, Onyeka and Omogbadegun, Zacchaeus and Aromolaran, Olufemi and Uwoghiren, Efosa and Olaniyan, Damilare and Olawole, Obembe},
  booktitle={2019 19th International Conference on Computational Science and Its Applications (ICCSA)}, 
  title={Data Clustering: Algorithms and Its Applications}, 
  year={2019},
  volume={},
  number={},
  pages={71-81},
  doi={10.1109/ICCSA.2019.000-1}}

@ARTICLE{decision,
  author={Wu, Tong and Liu, Xinwang and Qin, Jindong and Herrera, Francisco},
  journal={IEEE Transactions on Fuzzy Systems}, 
  title={Balance Dynamic Clustering Analysis and Consensus Reaching Process With Consensus Evolution Networks in Large-Scale Group Decision Making}, 
  year={2021},
  volume={29},
  number={2},
  pages={357-371},
  doi={10.1109/TFUZZ.2019.2953602}}

@InProceedings{decisionreview,
author="Caruso, Giulia
and Gattone, Stefano Antonio
and Fortuna, Francesca
and Di Battista, Tonio",
editor="Bucciarelli, Edgardo
and Chen, Shu-Heng
and Corchado, Juan M.",
title="Cluster Analysis as a Decision-Making Tool: A Methodological Review",
booktitle="Decision Economics: In the Tradition of Herbert A. Simon's Heritage",
year="2018",
publisher="Springer International Publishing",
address="Cham",
pages="48--55",
isbn="978-3-319-60882-2"
}

@article{Amaro_2023,
doi = {10.1088/1361-6501/acf402},
url = {https://dx.doi.org/10.1088/1361-6501/acf402},
year = {2023},
month = {sep},
publisher = {IOP Publishing},
volume = {34},
number = {12},
pages = {125024},
author = {F D Amaro and R Antonietti and E Baracchini and L Benussi and S Bianco and F Borra and C Capoccia and M Caponero and D S Cardoso and G Cavoto and I A Costa and G D’Imperio and E Danè and G Dho and F Di Giambattista and E Di Marco and F Iacoangeli and E Kemp and H P Lima Júnior and G S P Lopes and G Maccarrone and R D P Mano and R R Marcelo Gregorio and D J G Marques and G Mazzitelli and A G McLean and P Meloni and A Messina and C M B Monteiro and R A Nobrega and I F Pains and E Paoletti and L Passamonti and F Petrucci and S Piacentini and D Piccolo and D Pierluigi and D Pinci and A Prajapati and F Renga and R J d C Roque and F Rosatelli and A Russo and G Saviano and N J C Spooner and R Tesauro and S Tomassini and S Torelli and D Tozzi and J M F dos Santos},
title = {Directional iDBSCAN to detect cosmic-ray tracks for the CYGNO experiment},
journal = {Measurement Science and Technology}
}

@article{Rodenko_2019,
doi = {10.1088/1742-6596/1189/1/012009},
url = {https://dx.doi.org/10.1088/1742-6596/1189/1/012009},
year = {2019},
month = {mar},
publisher = {IOP Publishing},
volume = {1189},
number = {1},
pages = {012009},
author = {S A Rodenko and A G Mayorov and V V Malakhov and I K Troitskaya and on behalf ofPAMELA collaboration},
title = {Track reconstruction of antiprotons and antideuterons in the coordinate-sensitive calorimeter of PAMELA spectrometer using the Hough transform},
journal = {Journal of Physics: Conference Series}
}

@article{DALITZ2019159,
title = {Automatic trajectory recognition in Active Target Time Projection Chambers data by means of hierarchical clustering},
journal = {Computer Physics Communications},
volume = {235},
pages = {159-168},
year = {2019},
issn = {0010-4655},
doi = {https://doi.org/10.1016/j.cpc.2018.09.010},
url = {https://www.sciencedirect.com/science/article/pii/S0010465518303242},
author = {Christoph Dalitz and Yassid Ayyad and Jens Wilberg and Lukas Aymans and Daniel Bazin and Wolfgang Mittig}
}

@article{weiharrowthaler,
  title = {Quantum algorithms for jet clustering},
  author = {Wei, Annie Y. and Naik, Preksha and Harrow, Aram W. and Thaler, Jesse},
  journal = {Phys. Rev. D},
  volume = {101},
  issue = {9},
  pages = {094015},
  numpages = {20},
  year = {2020},
  month = {May},
  publisher = {American Physical Society},
  doi = {10.1103/PhysRevD.101.094015},
  url = {https://link.aps.org/doi/10.1103/PhysRevD.101.094015}
}

@article{kerenidis,
  title = {Quantum spectral clustering},
  author = {Kerenidis, Iordanis and Landman, Jonas},
  journal = {Phys. Rev. A},
  volume = {103},
  issue = {4},
  pages = {042415},
  numpages = {11},
  year = {2021},
  month = {Apr},
  publisher = {American Physical Society},
  doi = {10.1103/PhysRevA.103.042415},
  url = {https://link.aps.org/doi/10.1103/PhysRevA.103.042415}
}

@inproceedings{qmeans,
 author = {Kerenidis, Iordanis and Landman, Jonas and Luongo, Alessandro and Prakash, Anupam},
 booktitle = {Advances in Neural Information Processing Systems},
 editor = {H. Wallach and H. Larochelle and A. Beygelzimer and F. d\textquotesingle Alch\'{e}-Buc and E. Fox and R. Garnett},
 pages = {},
 publisher = {Curran Associates, Inc.},
 title = {q-means: A quantum algorithm for unsupervised machine learning},
 url = {https://proceedings.neurips.cc/paper_files/paper/2019/file/16026d60ff9b54410b3435b403afd226-Paper.pdf},
 volume = {32},
 year = {2019}
}

@article{marketresearch,
author = {Girish Punj and David W. Stewart},
title ={Cluster Analysis in Marketing Research: Review and Suggestions for Application},

journal = {Journal of Marketing Research},
volume = {20},
number = {2},
pages = {134-148},
year = {1983},
doi = {10.1177/002224378302000204},

URL = { 
    
        https://doi.org/10.1177/002224378302000204
    
    

},
eprint = { 
    
        https://doi.org/10.1177/002224378302000204
    
    

}
}

@article{HUANG2007313,
title = {Marketing segmentation using support vector clustering},
journal = {Expert Systems with Applications},
volume = {32},
number = {2},
pages = {313-317},
year = {2007},
issn = {0957-4174},
doi = {https://doi.org/10.1016/j.eswa.2005.11.028},
url = {https://www.sciencedirect.com/science/article/pii/S0957417405003404},
author = {Jih-Jeng Huang and Gwo-Hshiung Tzeng and Chorng-Shyong Ong}
}

@inproceedings{targetedads,
author = {Wu, Xiaohui and Yan, Jun and Liu, Ning and Yan, Shuicheng and Chen, Ying and Chen, Zheng},
title = {Probabilistic Latent Semantic User Segmentation for Behavioral Targeted Advertising},
year = {2009},
isbn = {9781605586717},
publisher = {Association for Computing Machinery},
address = {New York, NY, USA},
url = {https://doi.org/10.1145/1592748.1592751},
doi = {10.1145/1592748.1592751},
booktitle = {Proceedings of the Third International Workshop on Data Mining and Audience Intelligence for Advertising},
pages = {10–17},
numpages = {8},
location = {Paris, France},
series = {ADKDD '09}
}

@Article{Dutta2020,
author={Dutta, Pratik
and Saha, Sriparna
and Pai, Sanket
and Kumar, Aviral},
title={A Protein Interaction Information-based Generative Model for Enhancing Gene Clustering},
journal={Scientific Reports},
year={2020},
month={Jan},
day={20},
volume={10},
number={1},
pages={665},
issn={2045-2322},
doi={10.1038/s41598-020-57437-5},
url={https://doi.org/10.1038/s41598-020-57437-5}
}

@ARTICLE{geneclassification,
  author={Wai-Ho Au and Chan, K.C.C. and Wong, A.K.C. and Yang Wang},
  journal={IEEE/ACM Transactions on Computational Biology and Bioinformatics}, 
  title={Attribute clustering for grouping, selection, and classification of gene expression data}, 
  year={2005},
  volume={2},
  number={2},
  pages={83-101},
  doi={10.1109/TCBB.2005.17}}

@Article{Wang2010,
author={Wang, Jianxin
and Li, Min
and Deng, Youping
and Pan, Yi},
title={Recent advances in clustering methods for protein interaction networks},
journal={BMC Genomics},
year={2010},
month={Dec},
day={01},
volume={11},
number={3},
pages={S10},
issn={1471-2164},
doi={10.1186/1471-2164-11-S3-S10},
url={https://doi.org/10.1186/1471-2164-11-S3-S10}
}

@article{ensembleppinteraction,
    author = {Asur, Sitaram and Ucar, Duygu and Parthasarathy, Srinivasan},
    title = "{An ensemble framework for clustering protein–protein interaction networks}",
    journal = {Bioinformatics},
    volume = {23},
    number = {13},
    pages = {i29-i40},
    year = {2007},
    month = {07},
    issn = {1367-4803},
    doi = {10.1093/bioinformatics/btm212},
    url = {https://doi.org/10.1093/bioinformatics/btm212},
    eprint = {https://academic.oup.com/bioinformatics/article-pdf/23/13/i29/50715669/bioinformatics\_23\_13\_i29.pdf},
}

@article{speechclustering,
title = {A robust unsupervised pattern discovery and clustering of speech signals},
journal = {Pattern Recognition Letters},
volume = {116},
pages = {254-261},
year = {2018},
issn = {0167-8655},
doi = {https://doi.org/10.1016/j.patrec.2018.10.035},
url = {https://www.sciencedirect.com/science/article/pii/S016786551830864X},
author = {Kishore Kumar R and Lokendra Birla and Sreenivasa Rao K}
}

@ARTICLE{clusteringsegmentation,
  author={Coleman, G.B. and Andrews, H.C.},
  journal={Proceedings of the IEEE}, 
  title={Image segmentation by clustering}, 
  year={1979},
  volume={67},
  number={5},
  pages={773-785},
  doi={10.1109/PROC.1979.11327}}

@InProceedings{Chang_2017_ICCV,
author = {Chang, Jianlong and Wang, Lingfeng and Meng, Gaofeng and Xiang, Shiming and Pan, Chunhong},
title = {Deep Adaptive Image Clustering},
booktitle = {Proceedings of the IEEE International Conference on Computer Vision (ICCV)},
month = {Oct},
year = {2017}
}

@inproceedings{clusterrec,
author = {Shepitsen, Andriy and Gemmell, Jonathan and Mobasher, Bamshad and Burke, Robin},
title = {Personalized Recommendation in Social Tagging Systems Using Hierarchical Clustering},
year = {2008},
isbn = {9781605580937},
publisher = {Association for Computing Machinery},
address = {New York, NY, USA},
url = {https://doi.org/10.1145/1454008.1454048},
doi = {10.1145/1454008.1454048},
booktitle = {Proceedings of the 2008 ACM Conference on Recommender Systems},
pages = {259–266},
numpages = {8},
location = {Lausanne, Switzerland},
series = {RecSys '08}
}

@inproceedings{theontology,
author = {Schickel-Zuber, Vincent and Faltings, Boi},
title = {Using Hierarchical Clustering for Learning Theontologies Used in Recommendation Systems},
year = {2007},
isbn = {9781595936097},
publisher = {Association for Computing Machinery},
address = {New York, NY, USA},
url = {https://doi.org/10.1145/1281192.1281257},
doi = {10.1145/1281192.1281257},
booktitle = {Proceedings of the 13th ACM SIGKDD International Conference on Knowledge Discovery and Data Mining},
pages = {599–608},
numpages = {10},
location = {San Jose, California, USA},
series = {KDD '07}
}

@INPROCEEDINGS{watershed,
  author={Ng, H.P. and Ong, S.H. and Foong, K.W.C. and Goh, P.S. and Nowinski, W.L.},
  booktitle={2006 IEEE Southwest Symposium on Image Analysis and Interpretation}, 
  title={Medical Image Segmentation Using K-Means Clustering and Improved Watershed Algorithm}, 
  year={2006},
  volume={},
  number={},
  pages={61-65},
  doi={10.1109/SSIAI.2006.1633722}}

@article{GAFFEY2010564,
title = {Space weathering and the interpretation of asteroid reflectance spectra},
journal = {Icarus},
volume = {209},
number = {2},
pages = {564-574},
year = {2010},
issn = {0019-1035},
doi = {https://doi.org/10.1016/j.icarus.2010.05.006},
url = {https://www.sciencedirect.com/science/article/pii/S0019103510001879},
author = {Michael J. Gaffey}
}

@InProceedings{Gao_2021_CVPR,
    author    = {Gao, Angela F. and Rasmussen, Brandon and Kulits, Peter and Scheller, Eva L. and Greenberger, Rebecca and Ehlmann, Bethany L.},
    title     = {Generalized Unsupervised Clustering of Hyperspectral Images of Geological Targets in the Near Infrared},
    booktitle = {Proceedings of the IEEE/CVF Conference on Computer Vision and Pattern Recognition (CVPR) Workshops},
    month     = {June},
    year      = {2021},
    pages     = {4294-4303}
}

@Article{Magano2023,
author={Magano, Duarte
and Buffoni, Lorenzo
and Omar, Yasser},
title={Quantum density peak clustering},
journal={Quantum Machine Intelligence},
year={2023},
month={Feb},
day={01},
volume={5},
number={1},
pages={9},
issn={2524-4914},
doi={10.1007/s42484-022-00090-0},
url={https://doi.org/10.1007/s42484-022-00090-0}
}

@software{github1,
  author       = {Dhruv Gopalakrishnan and Luca Dellantonio and Antonio Di Pilato and Wahid Redheb and Felice Pantaleo and Michele Mosca},
  title        = {QLUE-algo/qlue: frontiers-paper},
  month        = jul,
  year         = 2024,
  publisher    = {Zenodo},
  version      = {frontiers-paper},
  doi          = {10.5281/zenodo.12655189},
  url          = {https://doi.org/10.5281/zenodo.12655189}
}

@book{Nielsen-Chuang-2010, 
place={Cambridge}, title={Quantum Computation and Quantum Information: 10th Anniversary Edition}, publisher={Cambridge University Press}, author={Nielsen, Michael A. and Chuang, Isaac L.}, year={2010}}

@misc{cmscollaboration2024review,
      title={Review of top quark mass measurements in CMS}, 
      author={CMS Collaboration},
      year={2024},
      eprint={2403.01313},
      archivePrefix={arXiv},
      primaryClass={hep-ex}
}

@thesis{Jaroslavceva:2865866,
      author        = "Jaroslavceva, Jekaterina",
      title         = "{A New Trackster Linking Algorithm Based on Graph Neural
                       Networks for the CMS Experiment at the Large Hadron
                       Collider at CERN}",
      school        = "Prague, Tech. U.",
      year          = "2023",
      url           = "https://cds.cern.ch/record/2865866",
      note          = "Presented 14 Jul 2023",
}

@article{CMS:2017jpq,
    collaboration = "CMS",
    title = "{The Phase-2 Upgrade of the CMS Endcap Calorimeter}",
    reportNumber = "CERN-LHCC-2017-023, CMS-TDR-019",
    doi = "10.17181/CERN.IV8M.1JY2",
    year = "2017"
}

@article{bioinfo,
author = {Jelili Oyelade and Itunuoluwa Isewon and Funke Oladipupo and Olufemi Aromolaran and Efosa Uwoghiren and Faridah Ameh and Moses Achas and Ezekiel Adebiyi},
title ={Clustering Algorithms: Their Application to Gene Expression Data},

journal = {Bioinformatics and Biology Insights},
volume = {10},
number = {},
pages = {BBI.S38316},
year = {2016},
doi = {10.4137/BBI.S38316},
    note ={PMID: 27932867},

URL = { 
    
        https://doi.org/10.4137/BBI.S38316
    
    

},
eprint = { 
    
        https://doi.org/10.4137/BBI.S38316
    
    

}
}

@article{Nash_2020,
doi = {10.1088/2058-9565/ab79b1},
url = {https://dx.doi.org/10.1088/2058-9565/ab79b1},
year = {2020},
month = {mar},
publisher = {IOP Publishing},
volume = {5},
number = {2},
pages = {025010},
author = {Beatrice Nash and Vlad Gheorghiu and Michele Mosca},
title = {Quantum circuit optimizations for NISQ architectures},
journal = {Quantum Science and Technology}
}

@inproceedings{rosenberg2007v,
  title={V-measure: A conditional entropy-based external cluster evaluation measure},
  author={Rosenberg, Andrew and Hirschberg, Julia},
  booktitle={Proceedings of the 2007 joint conference on empirical methods in natural language processing and computational natural language learning (EMNLP-CoNLL)},
  pages={410--420},
  year={2007}
}

@article{Haug_2023,
doi = {10.1088/2632-2153/acb0b4},
url = {https://dx.doi.org/10.1088/2632-2153/acb0b4},
year = {2023},
month = {jan},
publisher = {IOP Publishing},
volume = {4},
number = {1},
pages = {015005},
author = {Tobias Haug and Chris N Self and M S Kim},
title = {Quantum machine learning of large datasets using randomized measurements},
journal = {Machine Learning: Science and Technology}
}

@article{qml,
author = {Maria Schuld, Ilya Sinayskiy and Francesco Petruccione},
title = {An introduction to quantum machine learning},
journal = {Contemporary Physics},
volume = {56},
number = {2},
pages = {172-185},
year = {2015},
publisher = {Taylor & Francis},
doi = {10.1080/00107514.2014.964942},
URL = { 
       https://doi.org/10.1080/00107514.2014.964942
      },
eprint = { 
          https://doi.org/10.1080/00107514.2014.964942
         }
}
\end{document}